\documentclass[%
 twocolumn,
 amsmath,amssymb,
 aps,
prb,
]{revtex4-1}

\usepackage{graphicx}
\usepackage{dcolumn}
\usepackage{bm}
\usepackage{color,soul}
\usepackage{tcolorbox}
\usepackage{float}
\usepackage{hyperref}

\begin{document}

\preprint{APS/123-QED}

\title{Polarization-controlled selective excitation of Mie resonances of dielectric nanoparticle on a coated substrate}

\author{D.\,A. Pidgayko}
\email{d.pidgayko@metalab.ifmo.ru}
\author{Z.\,F. Sadrieva}
\author{K.\,S. Ladutenko}
\author{A.\,A. Bogdanov}
\email{a.bogdanov@metalab.ifmo.ru}
\affiliation{Department of Physics and Engineering, ITMO University, 197101, Saint-Petersburg, Russia
}

\date{\today}
 
\begin{abstract} Spherical high-index nanoparticles with low material losses
support sharp high-Q electric and magnetic resonances and exhibit a number
of interesting optical phenomena. Developments in fabrication techniques have
enabled further study of their properties and the investigation of related
optical effects. After deposition on a substrate, the optical properties of
a particle change dramatically due to mutual interaction. Here, we consider
a silicon spherical nanoparticle on a dielectric one-layered substrate. At
normal incidence of light, the layer thickness controls the contribution
of the nanoparticle's electric and magnetic multipoles to the subsequent
optical response. We show that changing the polarization of incident light
at a specific excitation angle and layer thickness leads to switching
between the multipoles. We further observe a related polarization-driven
control over the direction of the scattered radiation. \end{abstract}

\maketitle


\section{Introduction} Plasmonic and high-index dielectric
resonant nanoparticles are one of the key \textcolor{black}{building blocks} in modern
nanophotonics~\cite{krasnok2012all, giannini2011plasmonic}. They
offer efficient control over light at the nanoscale whether single or packed in an array, and have \textcolor{black}{a number of} applications~\cite{jahani2016all,
aieta2015multiwavelength, miroshnichenko2012fano, evlyukhin2010optical}.
For example, plasmonic nanoparticles exhibit strong field
localization~\cite{hutter2004exploitation,maier2007plasmonics}, and
are used for SERS~\cite{dornhaus1980surface,alvarez2007nanoimprinted},
sensing~\cite{melendez1996commercial, taylor2017single}, as well as for
chemical and biological applications~\cite{Novel,brolo2012plasmonics}.
However, the high intrinsic losses of plasmonic materials limit their
use for certain applications. High-index dielectric nanoparticles, on the
other hand, have significantly lower losses and support both electric and
magnetic responses \textcolor{black}{in the visible and near-IR regions}~\cite{kuznetsov2012magnetic, staude2013tailoring}. This
leads to several new phenomena such as Kerker effect~\cite{nieto2011angle,
liu2018generalized, alaee2015generalized}, \textcolor{black}{directional scattering~\cite{shamkhi2019transverse,lu2015directional,liu2018generalized} or excitation of surface}
plasmon-polariton~\cite{sinev2017chirality, sinev2020steering}. Finally, dielectric nanoparticles
have enabled many non-liner photonic effects~\cite{shcherbakov2014enhanced, makarov2015tuning,shcherbakov2015ultrafast,carletti2019high, koshelev2020subwavelength}, implementation of subwavelength room-temperature lasers \cite{tiguntseva2020room}.

The optical properties of spherical nanoparticles are described by Mie
theory~\cite{bohren2008absorption, kerker2013scattering}, where high-index
dielectric nanoparticle\textcolor{black}{s} exhibit pronounced optical resonances. \textcolor{black}{Their fundamental mode (with the lowest frequency) is a} magnetic dipole (MD) resonance, followed by an electric dipole (ED), a magnetic quadrupole (MQ), and so on into higher-order multipoles.
Mie theory predicts the spectral position and quality factor for each multipole
separately, but there is some overlap of multipoles and several multipoles
are excited at the same wavelength.
In some cases, selective excitation
of multipoles is necessary~\cite{van2013designing,koshelev2020subwavelength}, such as the selective
excitation of MD to enhance second~\cite{carletti2015enhanced} and third
harmonic generation~\cite{melik2017selective}. Another example is the enlarged
optical pulling and pushing forces~\cite{liu2017optical} caused by manipulation
of the ED, MD, and MQ of the dielectric nanoparticles. Finally, control over the
contribution of multipoles to scattering provides multicolor pixels at the
nanoscale~\cite{nagasaki2017all, flauraud2017silicon, xiang2018manipulating, li2019phase}.

There are a number of methods for the selective excitation of multipole
resonances of high-index particles. One of them is structured light in the
form of tightly focused cylindrical vector-beams~\cite{wozniak2015selective}.
Alternatively, radially polarized light~\cite{neugebauer2016polarization}
or even a simple plane wave excitation~\cite{xiang2018manipulating,
sinev2016polarization, van2013designing} are also possible.
\textcolor{black}{An efficient way to control multipoles was proposed by Xiang
et~al.~\cite{xiang2018manipulating}, where the authors suggested using the evanescent field for excitation by illuminating the nanoantenna from the substrate side under the total internal reflection condition. 
Here, we use far-field excitation, which is more practical.
Sinev et~al. in Ref.~\cite{sinev2016polarization} 
demonstrated the  polarization-controlled enhancement
of the MD of a silicon nanoparticle on a plasmonic substrate. Alternatively, Van de~Groep and
Polman~\cite{van2013designing} showed theoretically that a dielectric one-layered substrate can be used to control ED and MD contribution to the scattering. However, they only considered the case of normal incidence, where due to the selection rules, only modes with the azimuthal number $m=\pm1$ were present. In term of the group symmetry, one can say that  only the modes from $E_1$ irreducible representation were excited~\cite{gladyshev2020symmetry}.}

\textcolor{black}{Here, we use oblique incident light, which contains all azimuthal harmonics and, thus, excites the modes with, \textit{m}=$0,\pm 1,\pm 2,\text{etc}$. It gives an additional degree of freedom for selective multipole excitation and manipulation of their interference. We show that the angle of incidence and polarization of the excitation light combined with adjusting the silica spacer, which separates the nanoantenna from the silicon substrate provide a flexible control over the relative amplitudes of MD and ED, and their interference which governs the scattered field. The thickness of the SiO$_2$ spacer defines the phase shift between ED and MD, while the incident angle defines their relative amplitudes. Thus, we can strongly enhance or almost completely suppress the ED or MD component in the scattered field or maximize the scattering in certain directions. For example, the demonstrated enhancement of scattering to the directions of light source can be used for {\it retroreflectors} of subwavelength thickness~\cite{arbabi2017planar}.}          




\section{Results and discussion}

\begin{figure}[t]\centering
`\includegraphics[width = 0.8\linewidth ]{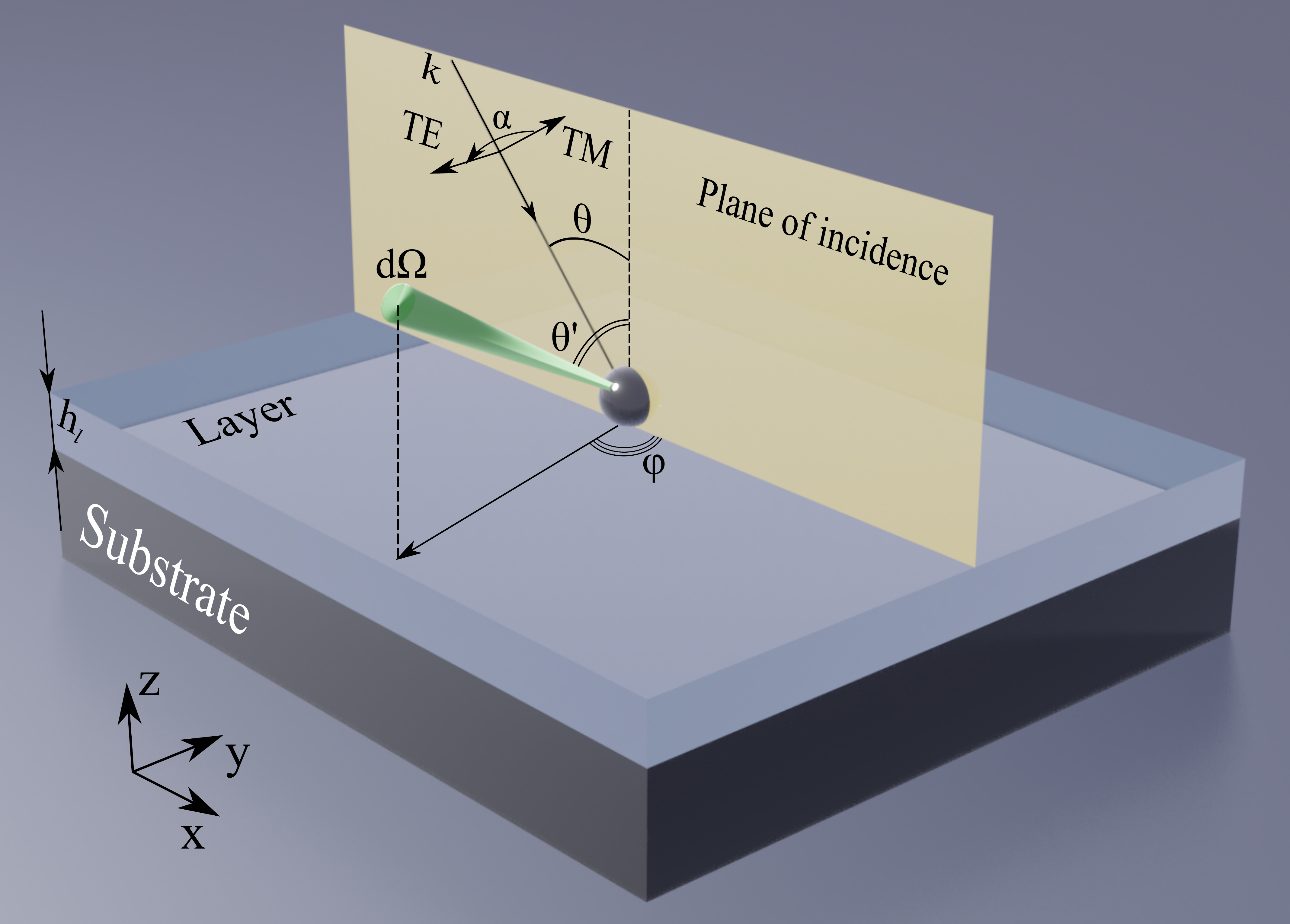}
\caption{ \label{Pic0} Geometry of the system. $\theta$ denotes angle of
incidence. $\theta'$ and $\varphi$ are stand for direction of scattered
radiation in a solid angle d$\Omega$. $\alpha$  indicates the direction of
polarization with values of 0$^{\circ}$ and 90$^{\circ}$ for TM and TE, respectively. }
\end{figure}

We consider scattering from a silicon spherical nanoparticle with fixed
radius $R = 85$~nm on top of a silica layer of variable thickness
backed by the silicon substrate (see Fig.~\ref{Pic0}). We employ the
T-matrix method for scattering from a single particle and the scattering
matrix method for propagation in a layered structure. These methods
have previously been implemented in open-source numerical software ``Smuthi''
using Python~\cite{egel2016efficient,egel2016light,egel2017extending}.
Simulation of the electromagnetic fields
inside the particle was performed in COMSOL Multiphysics package.

It is important to note that an analytical solution of the scattering problem for a small dielectric or plasmonic particle can be indeed obtained even for the case of multilayered substrate and oblique incident plane wave in a point-dipole approximation using the Green's function formalism \cite{miroshnichenko2015substrate}. We used this method, for example, in \cite{sinev2017chirality, sinev2020steering}. However, the analytical solutions are quite cumbersome for the analysis as it requires calculation of Sommerfeld's integral and solutions of a transcendent equation for complex poles of the leaky modes. Therefore, the numerical analysis is almost unavoidable. Moreover, in Sec. II.C we analyzed the directivity of the scattered field taking into account magnetic quadrupole resonance. These regimes are beyond the dipole approximation and also require numerical calculations for the correct analysis of scattering. Finally, the analytical solution in a point-dipole approximation contains a fitting parameter --- the height at which the point dipole should be positioned. The scattering from the nanoantenna strongly depends on this parameter (see additional materials to \cite{miroshnichenko2015substrate}).

\begin{figure}[t]\centering
	\includegraphics[width =1\linewidth ]{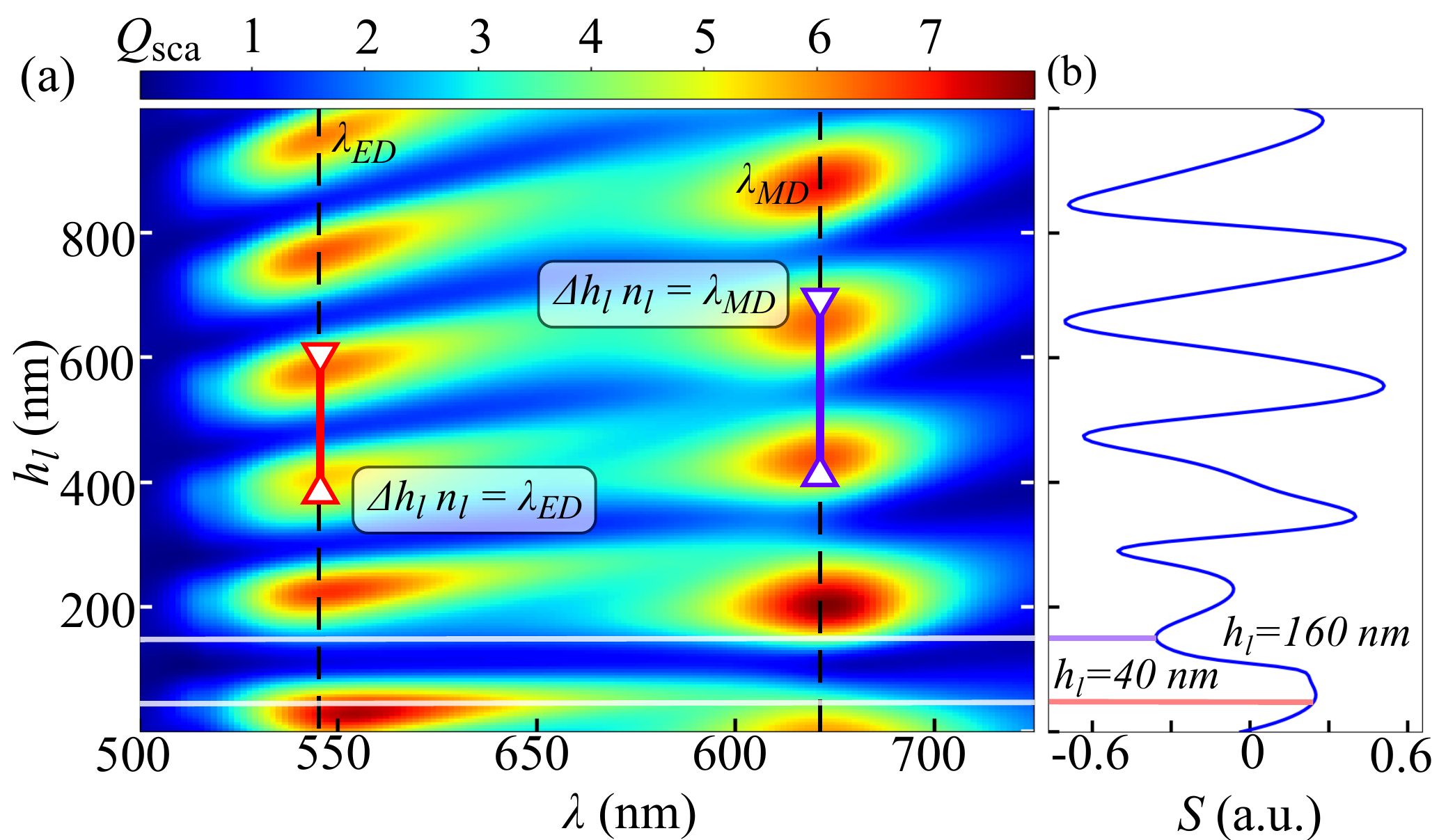}
	\caption{
		\label{Pic1}
	(a) Scattering efficiency map for normal incidence in $\lambda$--\textit{h}$_{l}$ axis.
	Dashed lines marks wavelengths of the ED and MD. White lines show layer thicknesses for special cases.
	(b)~Selectivity, calculated from the scattering efficiency map along the dashed lines. \textit{h}$_{l}$ axis is common with (a).
	Red and purple lines shows extreme values of selectivity with the ED/MD dominance in scattering, respectively.}
\end{figure}

\subsection{Normal incidence}

Firstly, varying the thickness of the layer \textit{h}$_{l}$,
we investigate the upper half-space scattering efficiency \textit{Q}$_{sca}$
at normal incidence, where scattering efficiency is a scattering
cross-section normalized to a geometrical cross-section.

\begin{figure*}[t]
	\includegraphics[width =0.9\linewidth ]{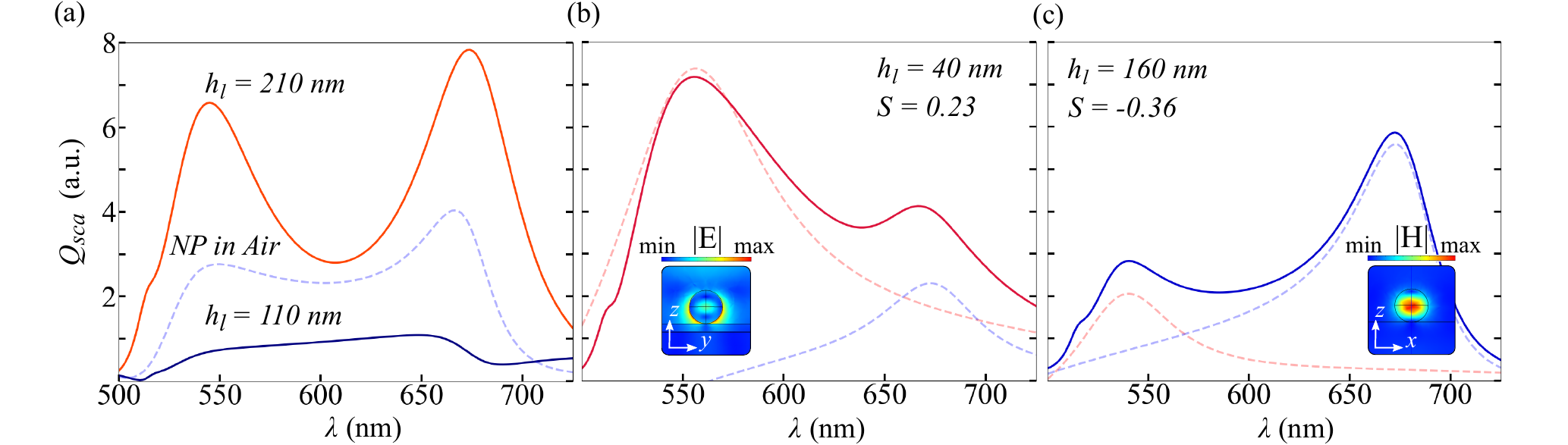}
	\caption{
		\label{Regimes}
Scattering efficiencies for different layer thicknesses. \textit{Q}$_{sca}$ axis is common for all panels. (a) Scattering for the simultaneous enhancement (orange) and suppression (dark blue) of the ED and MD. For comparison, \textit{Q}$_{sca}$ for the particle in free space is added. (b) and (c) Scattering efficiencies with the ED and MD domination regimes. Thin red and blue dashed lines shows Fano fitting curves for the ED and MD, respectively. Insets hows fields distribution at the corresponding resonances.}
\end{figure*}

At zero \textit{h}$_{l}$, with the particle located on a silicon substrate,
the scattering is enhanced in comparison with the free space case. In spite
the interaction with the substrate, there is no spectral shift of the ED and MD resonances
(see Appendix A, Fig. \ref{Apendix_a}). Infinite \textit{h}$_{l}$ corresponds to the
case of a particle on glass, where the scattering enhancement is negligible.
For intermediate cases, the layer modulates the standing wave
in the upper half-space, that results in scattering modulation (see
Fig.~\ref{Pic1}a). As MD (at 670 nm) and ED (at 540 nm) interact
with the standing wave, along the \textit{h}$_{l}$ axis we note
oscillating behavior of the \textit{Q}$_{sca}$ with a resonant condition:
\begin{equation}
    d_0 + h_{l} n_{l} = m\lambda,
\end{equation}
where \textit{n}$_{l}$ is a refractive index of the layer, \textit{m}
is an integer, and $d_0$ is associated with initial scattering phase.
The presence of the initial scattering phase, which is different for
the ED and MD, makes it possible to find layer thicknesses where
simultaneous enhancement or suppression of both dipoles occurs (see
Fig.~\ref{Regimes}a). Alternatively, dipole enhancement takes place
at different layer thicknesses. For example, enhancement
of the ED can be achieved  at \textit{h}$_{l} = 40$~nm, and of the MD
at \textit{h}$_{l} =160$~nm (see Fig.~\ref{Regimes}b and \ref{Regimes}c).

\begin{figure}[t]\centering
	\includegraphics[width =0.9\linewidth ]{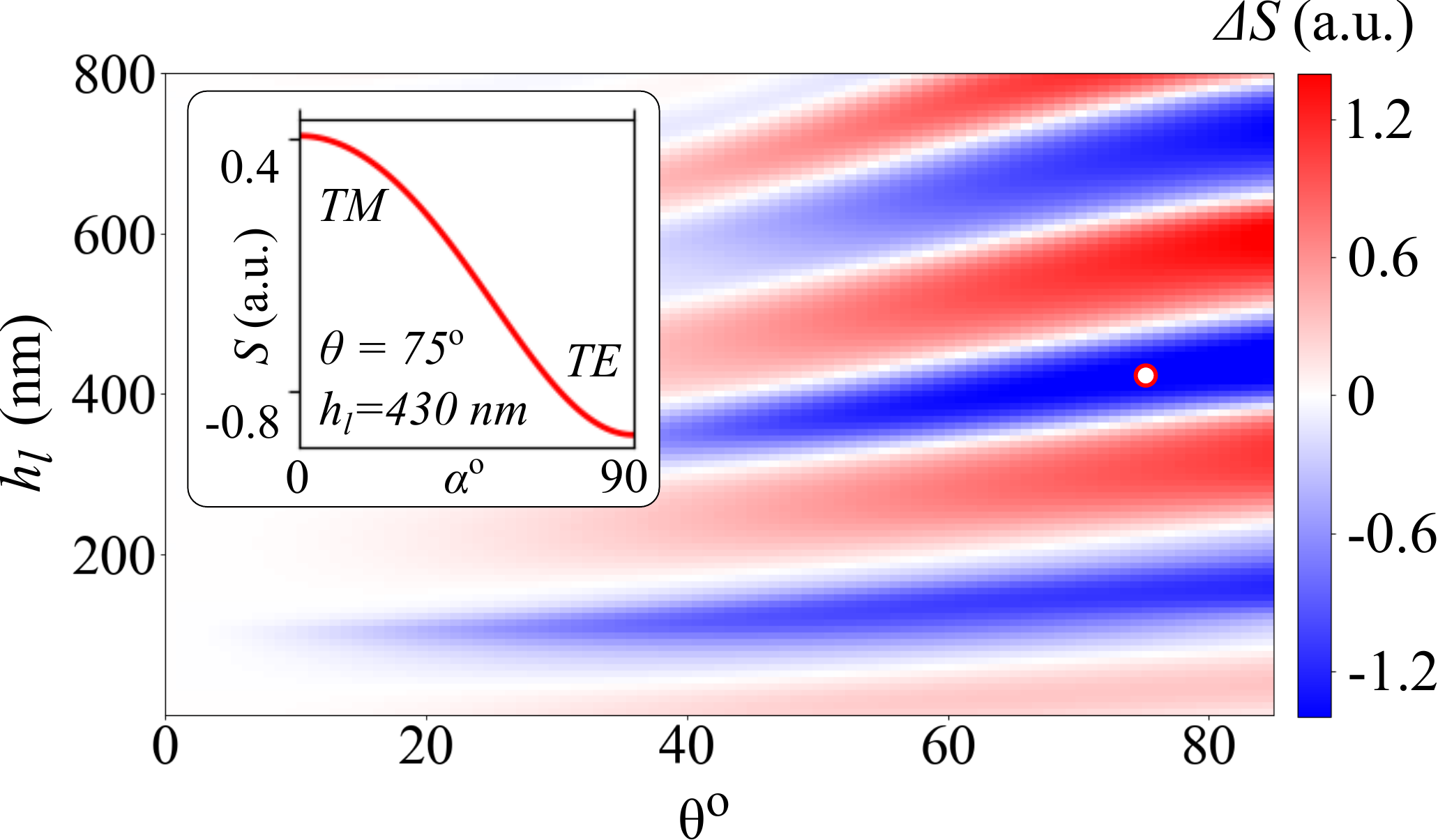}
	\caption{
		\label{Selectivity}
Selectivity variation for different angle of incidence and the layer thicknesses. A white dot shows, that for \mbox{\textit{h}$_{l}$= 430 nm}  and $\theta = 75^{\circ}$ a switch of polarisation from TE to TM leads to selectivity alter equals to -1.3. At these paremeters, inset shows the selectivity tuning when polarization angle $\alpha$ is rotated.}
\end{figure}

In order to quantify the relative contributions of
dipoles, we introduce selectivity, defined as:
\begin{equation}
    S = \frac{Q(\lambda_{ED})-Q(\lambda_{MD})}{Q(\lambda_{ED})+Q(\lambda_{MD})},
\end{equation}
where  $Q(\lambda_{ED})$ and $Q(\lambda_{MD})$ are scattering efficiencies
at wavelengths of the ED and MD resonances (see Fig.~\ref{Pic1}b). At
thicknesses of 40~nm and 160~nm, the selectivity is 0.23 and -0.36, respectively.
In these states, we provide the fit of \textit{Q}$_{sca}$ by two Fano-like
curves to represent the ED and MD contribution (thin red and blue dashed lines in
Figs.~\ref{Regimes}b and~\ref{Regimes}c). For example, at the wavelength of the enhanced
ED, the contribution of the suppressed MD to the scattering efficiency
is negligible (see Fig.~\ref{Regimes}b), the enhanced MD effect being similar. 
The dominance of one dipole over another also appears as a characteristic pattern in
the field distribution inside the particle (see inserts in Fig~\ref{Regimes}b and c).
It should be noted that the absolute value of the selectivity is relatively
small due to the overlapping of the
enhanced and suppressed
\begin{figure}[t]\centering
	\includegraphics[width =1\linewidth ]{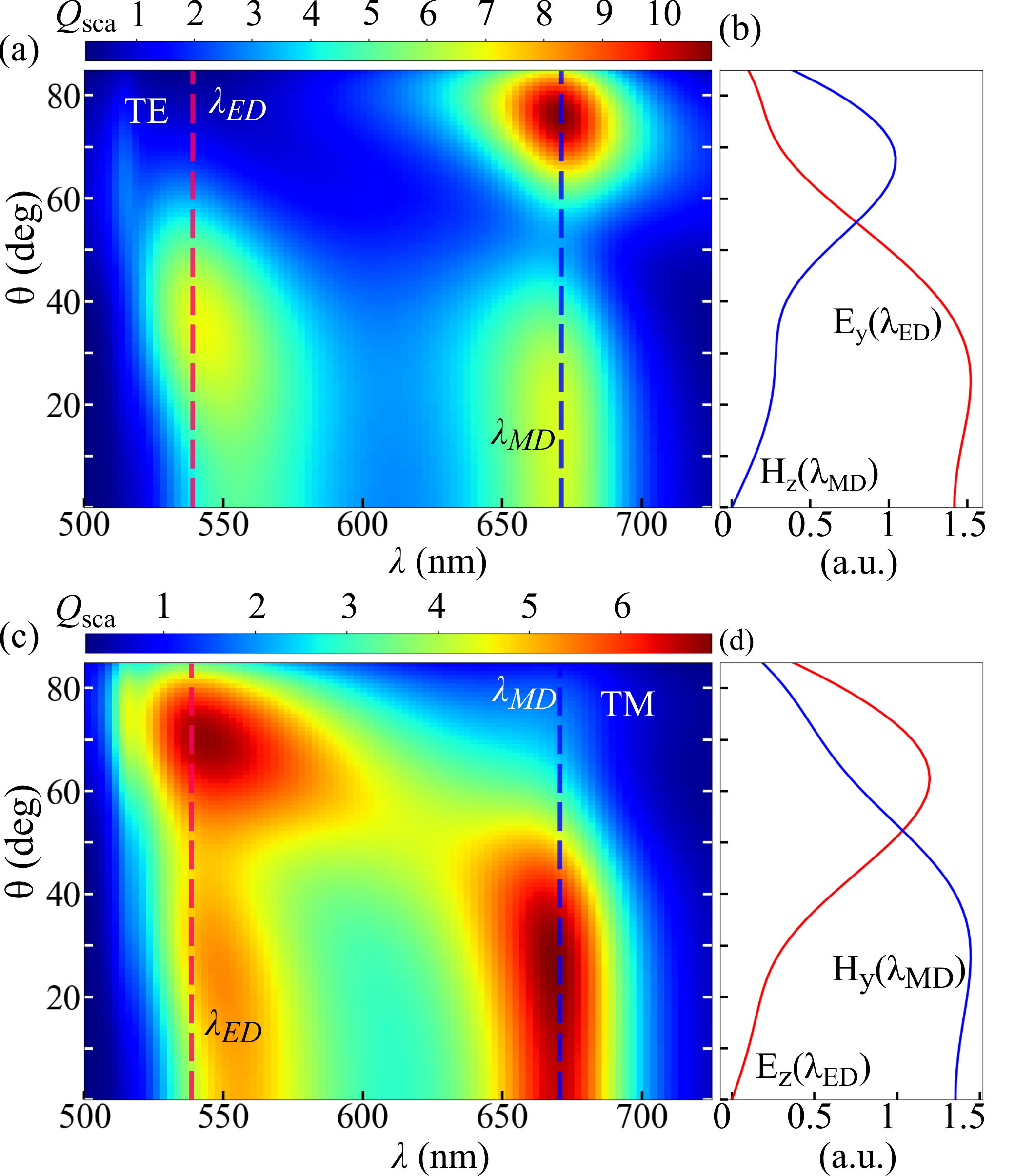}
	\caption{
		\label{Pic2}
Scattering efficiency maps as a function of the incidence angle for TE (a) and TM (c) polarization for \mbox{\textit{h}$_{l}$= 430 nm}. Normalized components of electric and magnetic fields for TE (b) and TM (d) polarization. Components of electric and magnetic fields are chosen at the wavelengths of the ED and the MD resonances respectively.}
\end{figure}
dipoles at the wavelength of the latter. As
we further increase the layer thickness, the selectivity oscillates, reaching
extreme values of 0.6 and -0.7 (see Fig.~\ref{Pic1}b). At the same time, these
values do not reflect the presence of the additional spectral features
associated with Fabri-Perot modes in thick layers. 
We have to therefore use additional degrees of freedom in order to
achieve a regime where only one dipole contributes to the scattering.

\subsection{Oblique incidence}

A single-dipole scattering regime can be achieved at an oblique incidence.
In this case, the polarization of the incident radiation begins to play a
significant role, allowing us to switch selectivity while keeping
the angle of incidence and the thickness of the layer constant. In
order to find the optimal values for the incidence angle, as well as the thickness of the
layer, we simulate the 

difference between selectivity for TE and TM polarization at different incidence angle and buffer layer thickness:
\begin{equation}
    \Delta  S = S_{TE}-S_{TM},
\end{equation}
where $S_{TE}$ and $S_{TM}$ are selectivity values for TE and TM incident
polarization.As far as the selectivity itself
\begin{figure}[t]\centering
	\includegraphics[width =0.65\linewidth ]{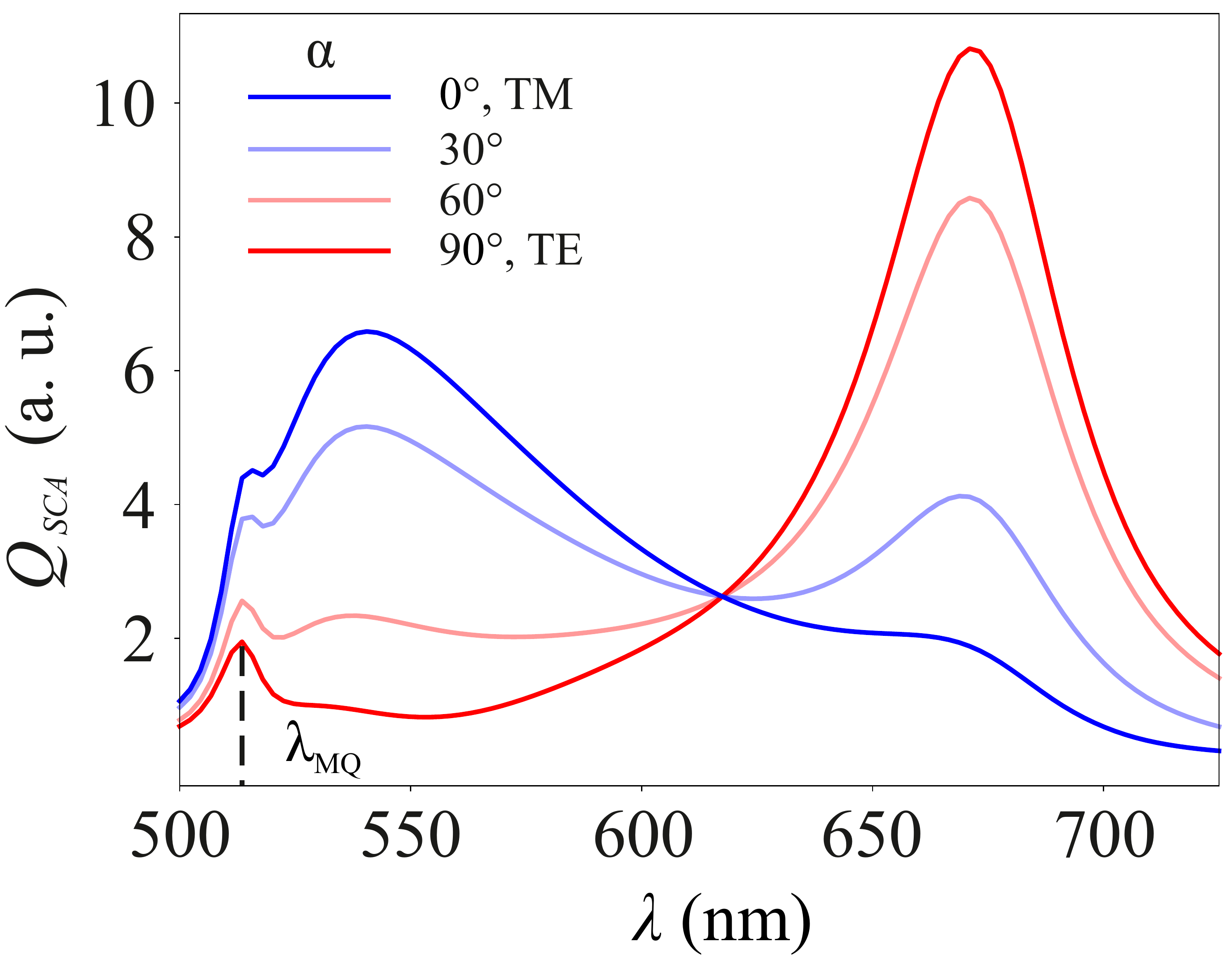}
	\caption{
		\label{Polarization switching}
Scattering spectra of Si nanoparticle on layer with \mbox{\textit{h}$_{l}$= 430 nm}  and angle of incidence $\theta = 75^{\circ}$ for different orientations of polarization $\alpha$. Dashed line represents wavelength of the MQ.}
\end{figure}
is an analog to the contrast between the contributions of electric and magnetic dipoles to scattering, the selectivity variation allows to determine the parameters that simultaneously achieve single-dipole scattering regimes and pronounced polarization-driven switching between them. Therefore, scattering efficiencies at ED and MD resonant wavelengths for both polarization are used for Fig.4.  There we see that the absolute value of
selectivity variation grows with the increase of the incidence angle,
while the dependence on the layer thickness is periodic in character.

Next, we demonstrate polarization switching of the selectivity with layer thickness equals 430~nm, and an incident angle of 75$^{\circ}$ (white dot in Fig.~\ref{Selectivity}).
Scattering efficiency maps of wavelength-incident angle coordinates for both
TE and TM polarization are shown in Figs.~\ref{Pic2}a and~\ref{Pic2}c.
At normal incidence, the ED and MD make approximately equal contribution to
scattering, which remains the same for both polarizations as the angle of
incidence increases. However, in the vicinity of 75$^{\circ}$, single-ED or
single-MD scattering regimes occur. Figure.~\ref{Pic2}a shows that, for
TE polarization, the enhanced MD almost completely dominates the suppressed
ED, a situation which is reversed with TM polarization (see Fig.~\ref{Pic2}c).

We explain this effect by considering the angular dependence of the background
field of the standing wave at wavelengths of the ED and MD in plane $z = R$
where $z = 0$ coincides with the upper boundary of the layer. We calculate
normalized fields using Fresnel coefficients then choose the determining
components of the fields. In the case of TE polarization (see Fig.~\ref{Pic2}b),
E$_y$($\lambda_{ED}$) has a maximum at 30$^{\circ}$ and then subsides, while
H$_z$($\lambda_{ED}$) reaches its maximum at 70$^{\circ}$. Since the magnitude
of the dipole is proportional to the applied field, the vertical magnetic dipole
provides most of the scattering, so the MD-only scattering regime is achieved.

Additionally, due to suppression of the ED it becomes possible to distinguish
the MQ contribution to scattering, dashed line in Fig.~\ref{Polarization
switching}, TE polarization. The ED-only scattering regime for TM polarization
occurs similarly.
Changing the polarization angle from TE to TM, we can
continuously tune the scattering regime from MD-only to ED-only (see
Fig.~\ref{Polarization switching} and inset in Fig.~\ref{Selectivity}).

\begin{figure}[t]
	\includegraphics[width =1\linewidth ]{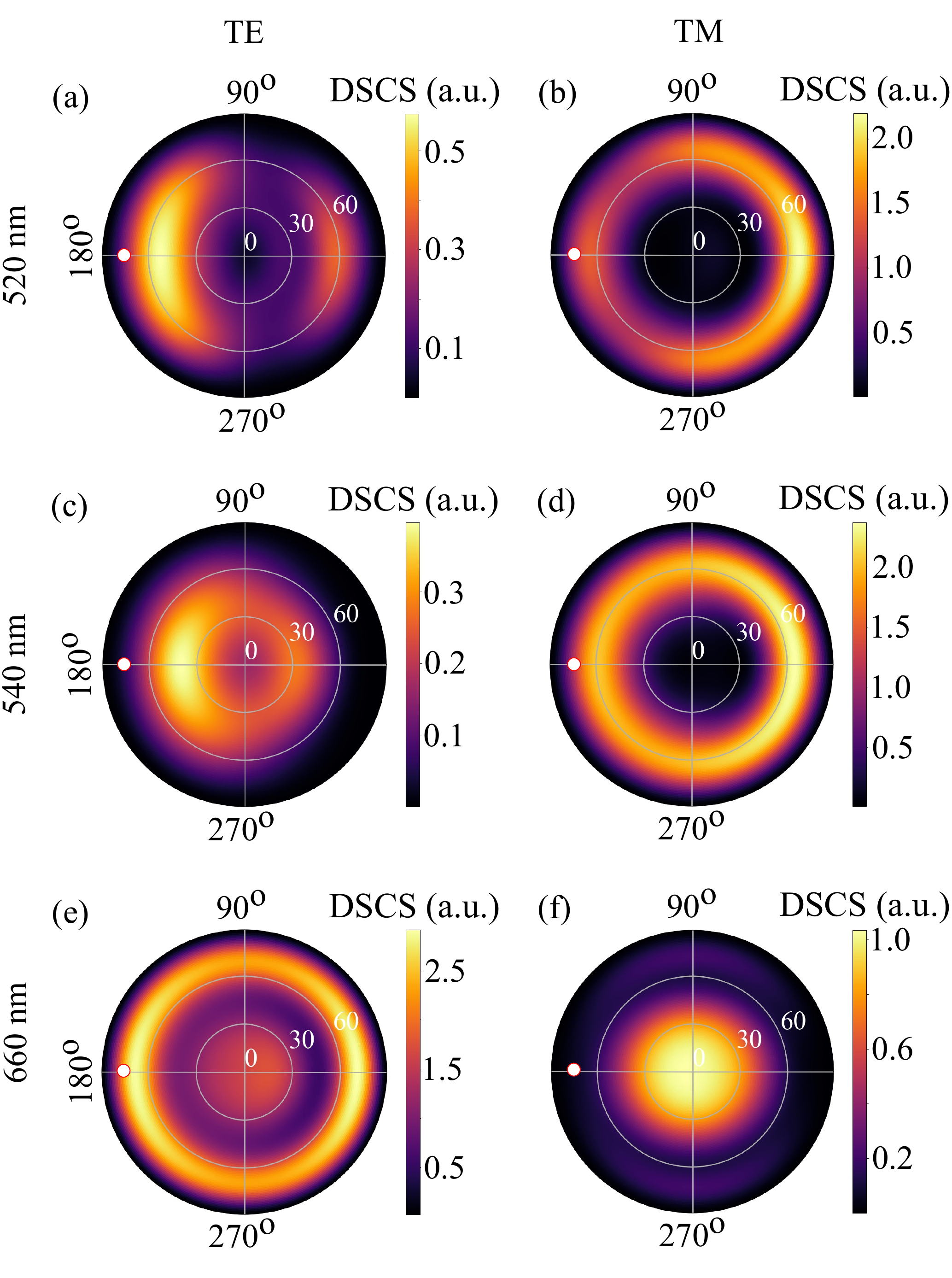}
	\caption{
		\label{Pic3}
Differential scattering cross sections for TE (a, c, e), and TM (b, d, f) polarization of the excitation and \mbox{\textit{h}$_{l}$= 430 nm},  and $\theta = 75^{\circ}$.  White dots indicate the direction of incident radiation.
	}
\end{figure}
\begin{figure}[h]
	\includegraphics[width =0.7\linewidth ]{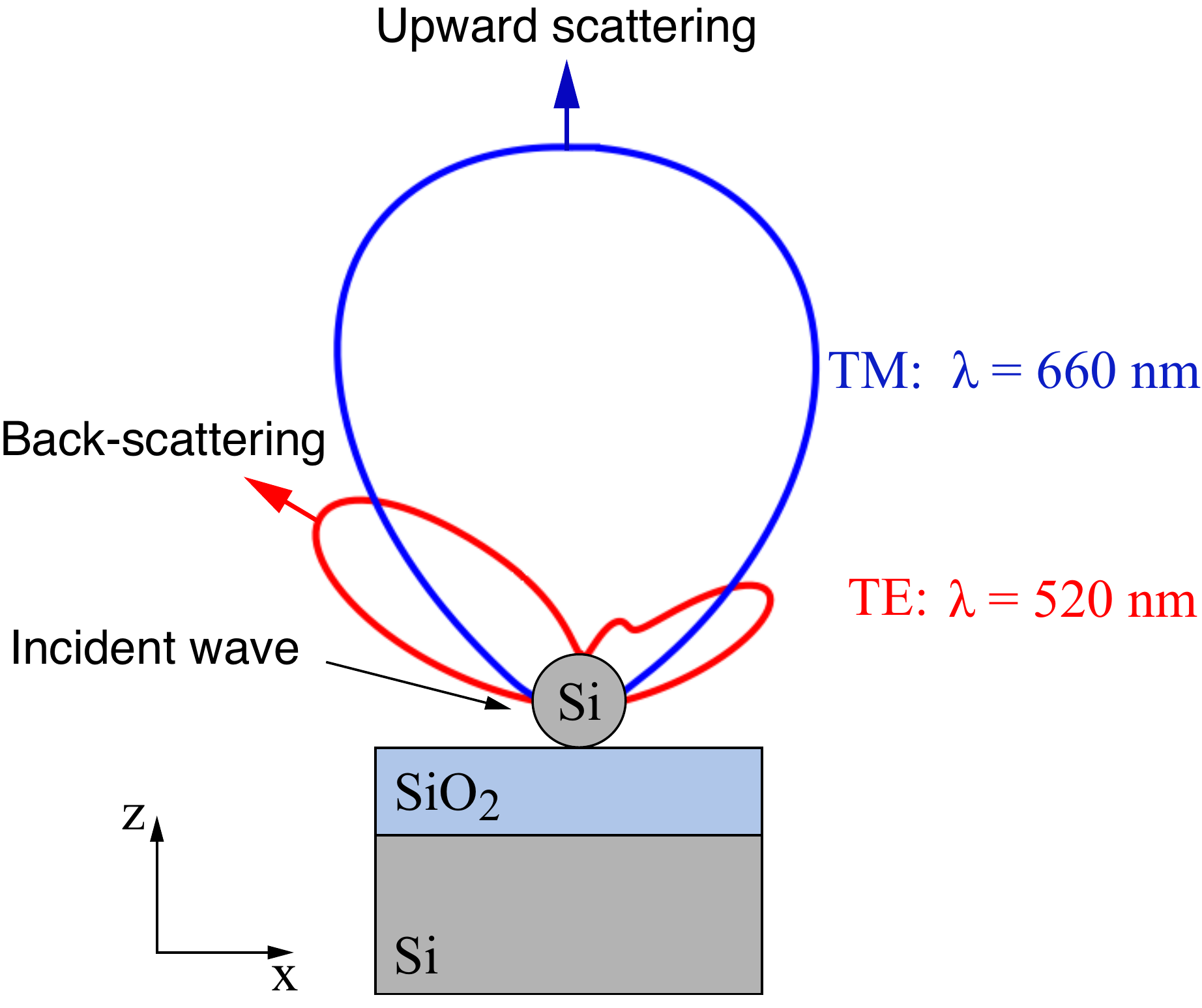}
	\caption{
		\label{Pic4}
Diagrams of the negative angle scattering (red curve) and the upward scatterings (blue curve) of Si nanoparticle on layer with \mbox{\textit{h}$_{l}$= 430 nm}  and angle of incidence $\theta = 75^{\circ}$. Data is taken from Figs.7a and 7f}.
\end{figure}
\subsection{Directivity of scattered radiation}

In the previous section we found an angle of incidence and layer thickness
that  enabled us tune the scattering regimes in the upper half-space. The
contribution of three main features, the MQ, MD, and ED, is controlled by
the polarization direction of the excitation. In this section, we investigate
the directivity of scattered radiation in peculiar regimes.  In the upper
half-space we simulate the differential scattering cross-section (DSCS)
normalized to the geometrical cross-section of the particle. The most
distinctive patterns are plotted in Fig.~\ref{Pic3} in polar coordinates. The
polar angle of the scattered radiation $\theta'$ is plotted along the radial
coordinate. The azimuthal angle $\varphi$ increases in a counterclockwise direction.
The direction of the incident radiation with \mbox{$\theta$ = 75$^{\circ}$} and
\mbox{$\varphi$ = 180$^{\circ}$} is indicated by the dots in Fig.~\ref{Pic3}.

The first regime of interest occurs in the vicinity of
the MQ, at $\lambda$ = 520 nm, TE polarization.
While the contribution of the ED is vastly suppressed  for TE-polarized
excitation, the MQ begins to noticeably affect the DSCS pattern.
Interference between radiation scattered by the ED and the MQ results in
negative-angle scattering (see Fig.~\ref{Pic3}a), which is plotted by a red curve
in Fig.~\ref{Pic4}, where the directivity pattern is plotted in the
plane of incidence ($xz$-plane). It should be noted that if we only consider ED and MD
in the simulation, negative-angle scattering disappears (see Appendix B, Fig. \ref{Appendix_b}). Switching the polarization to TM significantly increases the
contribution of the ED, and  a dipole-like DSCS with small
asymmetry is achieved (see Fig.~\ref{Pic3}b). The pattern becomes symmetric at the
wavelength of the ED resonance (see Fig.~\ref{Pic3}d), and similarly at
a wavelength of $\lambda_{MD}$, TE polarization
(see Fig.~\ref{Pic3}e). Switching polarization back to TM, we suppress
the MD and the contributions of both dipoles become comparable, leading to
directional upward scattering (see Fig.~\ref{Pic3}f and Fig.~\ref{Pic4}, blue curve).

\section{\label{Conclusion}Conclusion}

In this article we show that a one-layered substrate is an effective platform
for the manipulation of resonances of a dielectric nanoparticle. At normal
incidence, the thickness of the layer controls the enhancement and suppression
of the ED and the MD. At oblique incidence, it becomes possible to control
the contribution of the ED and the MD to the optical response through
polarization of the incident light. We futher present conditions where one
dipole resonance is almost completely suppressed as the other is enhanced, and
where a smooth transition to the reverse is also possible. Finally, we present
negative angle and upward scattering regimes, and show that adjusting the ED
and the MD contributions controls directivity of scattered radiation.

~\

\section{Acknowledgement}

This work is supported by RFBR, project number  18-29-20063. The multipole expansion calculations are supported by RFBR (19-02-00419). The Authors acknowledge Ian Dick for proofreading.

\section*{\label{Apendix_a} Appendix A: Comparison of scattering efficiencies for the nanoparticle on different substrates}

Here, we simulate scattering efficiency in the upper half space of a spherical silicon nanoparticle with the radius $R = 85$~nm, placed in the air, on silicon and glass substrate. On the silicon substrate the scattering is enhanced in comparison with the free space case with no spectral shift of the ED and MD resonances. For the particle on the glass scattering enhancement is negligible.
\begin{figure}[ht]\centering
	\includegraphics[width =1\linewidth ]{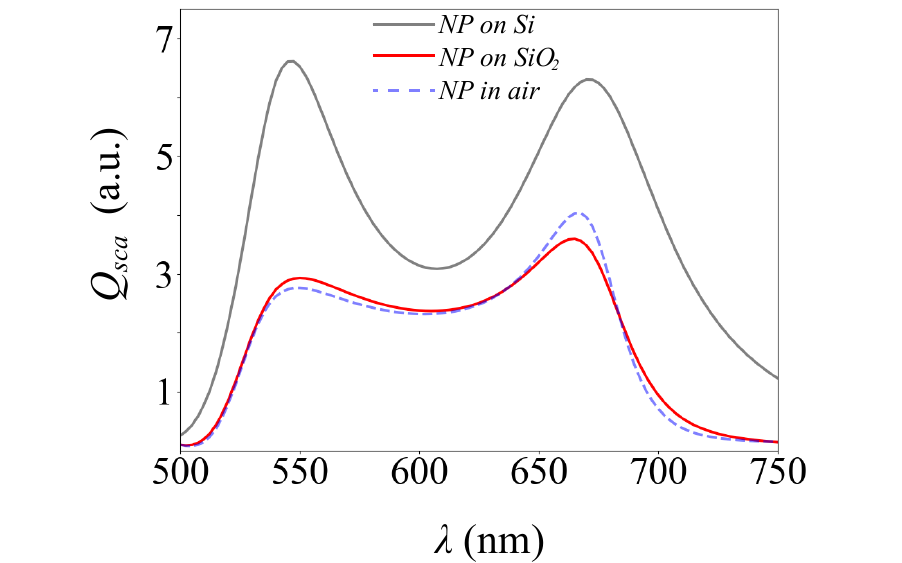}
	\caption{
		\label{Apendix_a}
	Scattering efficiency for the spherical silicon nanoparticle on different substartes.}
\end{figure}
\begin{figure*}[ht]\centering
	\includegraphics[width =0.8\linewidth]{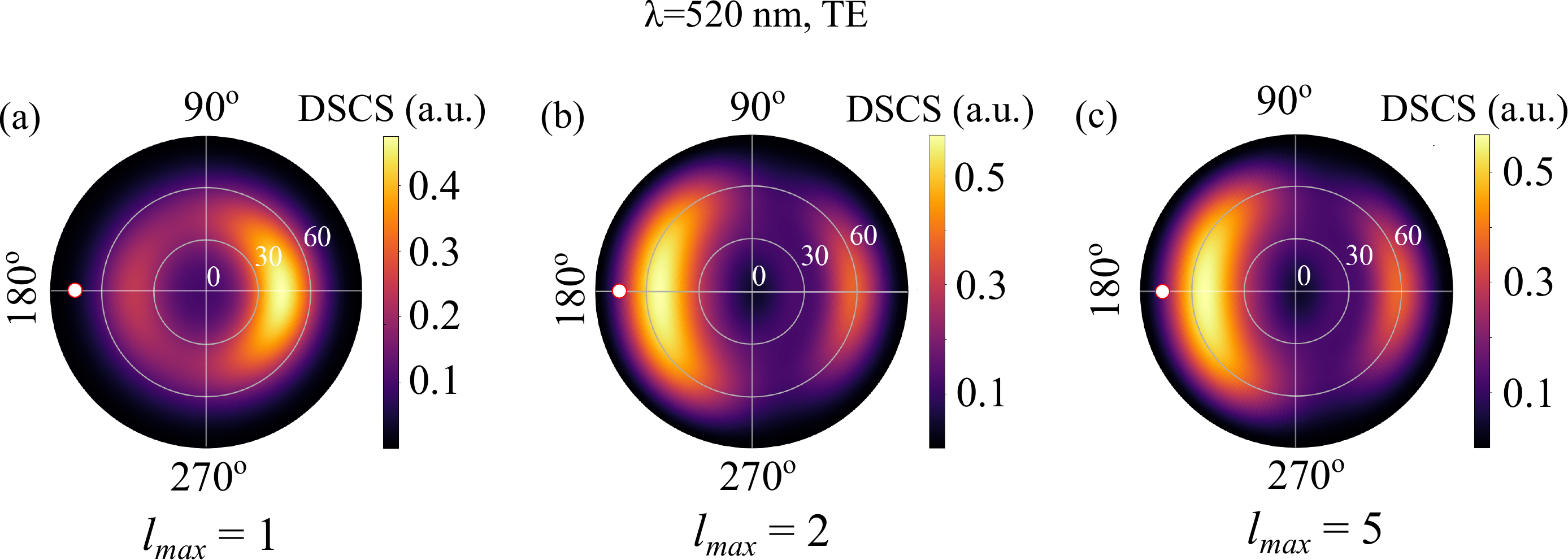}
	\caption{
		\label{Appendix_b}
	DSCS patterns for the spherical silicon nanoparticle, placed on the silicon substrate with the silica layer with the thickness of 430~nm. The particle is irradiated by TE polarized light at the wavelength $\lambda$=520~nm. One(a), two (b) and five (c) multipoles are taken into account for the simulation.}
\end{figure*}
\section*{\label{Apendix_b} Appendix B: Modification of the differential scattering cross section pattern}

In order to show, that the negative angle scattering occurs due to the interference between radiation scattered by the ED and the MQ, we simulated DSCS for the silicon nanoparticle on one-layered substrate, with the parameters given in the article, taking into account different maximal order of multipoles. 
In S 2(a) it is seen, that considering only dipole terms, we don't have negative angle scattering. Taking into account quadrupole, we obtain negative angle scattering. Further increase of the maximal order of multipoles, the DSCS pattern doesn't change.

\newpage
\bibliography{Main.bib}

\begin{thebibliography}{48}%
\makeatletter
\providecommand \@ifxundefined [1]{%
 \@ifx{#1\undefined}
}%
\providecommand \@ifnum [1]{%
 \ifnum #1\expandafter \@firstoftwo
 \else \expandafter \@secondoftwo
 \fi
}%
\providecommand \@ifx [1]{%
 \ifx #1\expandafter \@firstoftwo
 \else \expandafter \@secondoftwo
 \fi
}%
\providecommand \natexlab [1]{#1}%
\providecommand \enquote  [1]{``#1''}%
\providecommand \bibnamefont  [1]{#1}%
\providecommand \bibfnamefont [1]{#1}%
\providecommand \citenamefont [1]{#1}%
\providecommand \href@noop [0]{\@secondoftwo}%
\providecommand \href [0]{\begingroup \@sanitize@url \@href}%
\providecommand \@href[1]{\@@startlink{#1}\@@href}%
\providecommand \@@href[1]{\endgroup#1\@@endlink}%
\providecommand \@sanitize@url [0]{\catcode `\\12\catcode `\$12\catcode
  `\&12\catcode `\#12\catcode `\^12\catcode `\_12\catcode `\%12\relax}%
\providecommand \@@startlink[1]{}%
\providecommand \@@endlink[0]{}%
\providecommand \url  [0]{\begingroup\@sanitize@url \@url }%
\providecommand \@url [1]{\endgroup\@href {#1}{\urlprefix }}%
\providecommand \urlprefix  [0]{URL }%
\providecommand \Eprint [0]{\href }%
\providecommand \doibase [0]{http://dx.doi.org/}%
\providecommand \selectlanguage [0]{\@gobble}%
\providecommand \bibinfo  [0]{\@secondoftwo}%
\providecommand \bibfield  [0]{\@secondoftwo}%
\providecommand \translation [1]{[#1]}%
\providecommand \BibitemOpen [0]{}%
\providecommand \bibitemStop [0]{}%
\providecommand \bibitemNoStop [0]{.\EOS\space}%
\providecommand \EOS [0]{\spacefactor3000\relax}%
\providecommand \BibitemShut  [1]{\csname bibitem#1\endcsname}%
\let\auto@bib@innerbib\@empty
\bibitem [{\citenamefont {Krasnok}\ \emph {et~al.}(2012)\citenamefont
  {Krasnok}, \citenamefont {Miroshnichenko}, \citenamefont {Belov},\ and\
  \citenamefont {Kivshar}}]{krasnok2012all}%
  \BibitemOpen
  \bibfield  {author} {\bibinfo {author} {\bibfnamefont {A.~E.}\ \bibnamefont
  {Krasnok}}, \bibinfo {author} {\bibfnamefont {A.~E.}\ \bibnamefont
  {Miroshnichenko}}, \bibinfo {author} {\bibfnamefont {P.~A.}\ \bibnamefont
  {Belov}}, \ and\ \bibinfo {author} {\bibfnamefont {Y.~S.}\ \bibnamefont
  {Kivshar}},\ }\href@noop {} {\bibfield  {journal} {\bibinfo  {journal} {Opt.
  Express}\ }\textbf {\bibinfo {volume} {20}},\ \bibinfo {pages} {20599}
  (\bibinfo {year} {2012})}\BibitemShut {NoStop}%
\bibitem [{\citenamefont {Giannini}\ \emph {et~al.}(2011)\citenamefont
  {Giannini}, \citenamefont {Fern{\'a}ndez-Dom{\'\i}nguez}, \citenamefont
  {Heck},\ and\ \citenamefont {Maier}}]{giannini2011plasmonic}%
  \BibitemOpen
  \bibfield  {author} {\bibinfo {author} {\bibfnamefont {V.}~\bibnamefont
  {Giannini}}, \bibinfo {author} {\bibfnamefont {A.~I.}\ \bibnamefont
  {Fern{\'a}ndez-Dom{\'\i}nguez}}, \bibinfo {author} {\bibfnamefont {S.~C.}\
  \bibnamefont {Heck}}, \ and\ \bibinfo {author} {\bibfnamefont {S.~A.}\
  \bibnamefont {Maier}},\ }\href@noop {} {\bibfield  {journal} {\bibinfo
  {journal} {Chem. Rev.}\ }\textbf {\bibinfo {volume} {111}},\ \bibinfo {pages}
  {3888} (\bibinfo {year} {2011})}\BibitemShut {NoStop}%
\bibitem [{\citenamefont {Jahani}\ and\ \citenamefont
  {Jacob}(2016)}]{jahani2016all}%
  \BibitemOpen
  \bibfield  {author} {\bibinfo {author} {\bibfnamefont {S.}~\bibnamefont
  {Jahani}}\ and\ \bibinfo {author} {\bibfnamefont {Z.}~\bibnamefont {Jacob}},\
  }\href@noop {} {\bibfield  {journal} {\bibinfo  {journal} {Nat.
  Nanotechnol.}\ }\textbf {\bibinfo {volume} {11}},\ \bibinfo {pages} {23}
  (\bibinfo {year} {2016})}\BibitemShut {NoStop}%
\bibitem [{\citenamefont {Aieta}\ \emph {et~al.}(2015)\citenamefont {Aieta},
  \citenamefont {Kats}, \citenamefont {Genevet},\ and\ \citenamefont
  {Capasso}}]{aieta2015multiwavelength}%
  \BibitemOpen
  \bibfield  {author} {\bibinfo {author} {\bibfnamefont {F.}~\bibnamefont
  {Aieta}}, \bibinfo {author} {\bibfnamefont {M.~A.}\ \bibnamefont {Kats}},
  \bibinfo {author} {\bibfnamefont {P.}~\bibnamefont {Genevet}}, \ and\
  \bibinfo {author} {\bibfnamefont {F.}~\bibnamefont {Capasso}},\ }\href@noop
  {} {\bibfield  {journal} {\bibinfo  {journal} {Science}\ }\textbf {\bibinfo
  {volume} {347}},\ \bibinfo {pages} {1342} (\bibinfo {year}
  {2015})}\BibitemShut {NoStop}%
\bibitem [{\citenamefont {Miroshnichenko}\ and\ \citenamefont
  {Kivshar}(2012)}]{miroshnichenko2012fano}%
  \BibitemOpen
  \bibfield  {author} {\bibinfo {author} {\bibfnamefont {A.~E.}\ \bibnamefont
  {Miroshnichenko}}\ and\ \bibinfo {author} {\bibfnamefont {Y.~S.}\
  \bibnamefont {Kivshar}},\ }\href@noop {} {\bibfield  {journal} {\bibinfo
  {journal} {Nano Lett.}\ }\textbf {\bibinfo {volume} {12}},\ \bibinfo {pages}
  {6459} (\bibinfo {year} {2012})}\BibitemShut {NoStop}%
\bibitem [{\citenamefont {Evlyukhin}\ \emph {et~al.}(2010)\citenamefont
  {Evlyukhin}, \citenamefont {Reinhardt}, \citenamefont {Seidel}, \citenamefont
  {Luk’yanchuk},\ and\ \citenamefont {Chichkov}}]{evlyukhin2010optical}%
  \BibitemOpen
  \bibfield  {author} {\bibinfo {author} {\bibfnamefont {A.~B.}\ \bibnamefont
  {Evlyukhin}}, \bibinfo {author} {\bibfnamefont {C.}~\bibnamefont
  {Reinhardt}}, \bibinfo {author} {\bibfnamefont {A.}~\bibnamefont {Seidel}},
  \bibinfo {author} {\bibfnamefont {B.~S.}\ \bibnamefont {Luk’yanchuk}}, \
  and\ \bibinfo {author} {\bibfnamefont {B.~N.}\ \bibnamefont {Chichkov}},\
  }\href@noop {} {\bibfield  {journal} {\bibinfo  {journal} {Phys. Rev. B}\
  }\textbf {\bibinfo {volume} {82}},\ \bibinfo {pages} {045404} (\bibinfo
  {year} {2010})}\BibitemShut {NoStop}%
\bibitem [{\citenamefont {Hutter}\ and\ \citenamefont
  {Fendler}(2004)}]{hutter2004exploitation}%
  \BibitemOpen
  \bibfield  {author} {\bibinfo {author} {\bibfnamefont {E.}~\bibnamefont
  {Hutter}}\ and\ \bibinfo {author} {\bibfnamefont {J.~H.}\ \bibnamefont
  {Fendler}},\ }\href@noop {} {\bibfield  {journal} {\bibinfo  {journal} {Adv.
  Mater.}\ }\textbf {\bibinfo {volume} {16}},\ \bibinfo {pages} {1685}
  (\bibinfo {year} {2004})}\BibitemShut {NoStop}%
\bibitem [{\citenamefont {Maier}(2007)}]{maier2007plasmonics}%
  \BibitemOpen
  \bibfield  {author} {\bibinfo {author} {\bibfnamefont {S.~A.}\ \bibnamefont
  {Maier}},\ }\href@noop {} {\emph {\bibinfo {title} {Plasmonics: fundamentals
  and applications}}}\ (\bibinfo  {publisher} {Springer Science \& Business
  Media},\ \bibinfo {year} {2007})\BibitemShut {NoStop}%
\bibitem [{\citenamefont {Dornhaus}\ \emph {et~al.}(1980)\citenamefont
  {Dornhaus}, \citenamefont {Benner}, \citenamefont {Chang},\ and\
  \citenamefont {Chabay}}]{dornhaus1980surface}%
  \BibitemOpen
  \bibfield  {author} {\bibinfo {author} {\bibfnamefont {R.}~\bibnamefont
  {Dornhaus}}, \bibinfo {author} {\bibfnamefont {R.~E.}\ \bibnamefont
  {Benner}}, \bibinfo {author} {\bibfnamefont {R.~K.}\ \bibnamefont {Chang}}, \
  and\ \bibinfo {author} {\bibfnamefont {I.}~\bibnamefont {Chabay}},\
  }\href@noop {} {\bibfield  {journal} {\bibinfo  {journal} {Surf. Sci.}\
  }\textbf {\bibinfo {volume} {101}},\ \bibinfo {pages} {367} (\bibinfo {year}
  {1980})}\BibitemShut {NoStop}%
\bibitem [{\citenamefont {Alvarez-Puebla}\ \emph {et~al.}(2007)\citenamefont
  {Alvarez-Puebla}, \citenamefont {Cui}, \citenamefont {Bravo-Vasquez},
  \citenamefont {Veres},\ and\ \citenamefont
  {Fenniri}}]{alvarez2007nanoimprinted}%
  \BibitemOpen
  \bibfield  {author} {\bibinfo {author} {\bibfnamefont {R.}~\bibnamefont
  {Alvarez-Puebla}}, \bibinfo {author} {\bibfnamefont {B.}~\bibnamefont {Cui}},
  \bibinfo {author} {\bibfnamefont {J.-P.}\ \bibnamefont {Bravo-Vasquez}},
  \bibinfo {author} {\bibfnamefont {T.}~\bibnamefont {Veres}}, \ and\ \bibinfo
  {author} {\bibfnamefont {H.}~\bibnamefont {Fenniri}},\ }\href@noop {}
  {\bibfield  {journal} {\bibinfo  {journal} {J. Phys. Chem. C}\ }\textbf
  {\bibinfo {volume} {111}},\ \bibinfo {pages} {6720} (\bibinfo {year}
  {2007})}\BibitemShut {NoStop}%
\bibitem [{\citenamefont {Melendez}\ \emph {et~al.}(1996)\citenamefont
  {Melendez}, \citenamefont {Carr}, \citenamefont {Bartholomew}, \citenamefont
  {Kukanskis}, \citenamefont {Elkind}, \citenamefont {Yee}, \citenamefont
  {Furlong},\ and\ \citenamefont {Woodbury}}]{melendez1996commercial}%
  \BibitemOpen
  \bibfield  {author} {\bibinfo {author} {\bibfnamefont {J.}~\bibnamefont
  {Melendez}}, \bibinfo {author} {\bibfnamefont {R.}~\bibnamefont {Carr}},
  \bibinfo {author} {\bibfnamefont {D.~U.}\ \bibnamefont {Bartholomew}},
  \bibinfo {author} {\bibfnamefont {K.}~\bibnamefont {Kukanskis}}, \bibinfo
  {author} {\bibfnamefont {J.}~\bibnamefont {Elkind}}, \bibinfo {author}
  {\bibfnamefont {S.}~\bibnamefont {Yee}}, \bibinfo {author} {\bibfnamefont
  {C.}~\bibnamefont {Furlong}}, \ and\ \bibinfo {author} {\bibfnamefont
  {R.}~\bibnamefont {Woodbury}},\ }\href@noop {} {\bibfield  {journal}
  {\bibinfo  {journal} {Sens. Actuator B-Chem.}\ }\textbf {\bibinfo {volume}
  {35}},\ \bibinfo {pages} {212} (\bibinfo {year} {1996})}\BibitemShut
  {NoStop}%
\bibitem [{\citenamefont {Taylor}\ and\ \citenamefont
  {Zijlstra}(2017)}]{taylor2017single}%
  \BibitemOpen
  \bibfield  {author} {\bibinfo {author} {\bibfnamefont {A.~B.}\ \bibnamefont
  {Taylor}}\ and\ \bibinfo {author} {\bibfnamefont {P.}~\bibnamefont
  {Zijlstra}},\ }\href@noop {} {\bibfield  {journal} {\bibinfo  {journal} {ACS
  Sensors}\ }\textbf {\bibinfo {volume} {2}},\ \bibinfo {pages} {1103}
  (\bibinfo {year} {2017})}\BibitemShut {NoStop}%
\bibitem [{\citenamefont {Meyerbr{\"o}ker}\ \emph {et~al.}(2013)\citenamefont
  {Meyerbr{\"o}ker}, \citenamefont {Kriesche},\ and\ \citenamefont
  {Zharnikov}}]{Novel}%
  \BibitemOpen
  \bibfield  {author} {\bibinfo {author} {\bibfnamefont {N.}~\bibnamefont
  {Meyerbr{\"o}ker}}, \bibinfo {author} {\bibfnamefont {T.}~\bibnamefont
  {Kriesche}}, \ and\ \bibinfo {author} {\bibfnamefont {M.}~\bibnamefont
  {Zharnikov}},\ }\href@noop {} {\bibfield  {journal} {\bibinfo  {journal} {ACS
  Appl. Mater. Interfaces}\ }\textbf {\bibinfo {volume} {5}},\ \bibinfo {pages}
  {2641} (\bibinfo {year} {2013})}\BibitemShut {NoStop}%
\bibitem [{\citenamefont {Brolo}(2012)}]{brolo2012plasmonics}%
  \BibitemOpen
  \bibfield  {author} {\bibinfo {author} {\bibfnamefont {A.~G.}\ \bibnamefont
  {Brolo}},\ }\href@noop {} {\bibfield  {journal} {\bibinfo  {journal} {Nat.
  Photonics}\ }\textbf {\bibinfo {volume} {6}},\ \bibinfo {pages} {709}
  (\bibinfo {year} {2012})}\BibitemShut {NoStop}%
\bibitem [{\citenamefont {Kuznetsov}\ \emph {et~al.}(2012)\citenamefont
  {Kuznetsov}, \citenamefont {Miroshnichenko}, \citenamefont {Fu},
  \citenamefont {Zhang},\ and\ \citenamefont
  {Luk’Yanchuk}}]{kuznetsov2012magnetic}%
  \BibitemOpen
  \bibfield  {author} {\bibinfo {author} {\bibfnamefont {A.~I.}\ \bibnamefont
  {Kuznetsov}}, \bibinfo {author} {\bibfnamefont {A.~E.}\ \bibnamefont
  {Miroshnichenko}}, \bibinfo {author} {\bibfnamefont {Y.~H.}\ \bibnamefont
  {Fu}}, \bibinfo {author} {\bibfnamefont {J.}~\bibnamefont {Zhang}}, \ and\
  \bibinfo {author} {\bibfnamefont {B.}~\bibnamefont {Luk’Yanchuk}},\
  }\href@noop {} {\bibfield  {journal} {\bibinfo  {journal} {Sci. Rep.}\
  }\textbf {\bibinfo {volume} {2}},\ \bibinfo {pages} {492} (\bibinfo {year}
  {2012})}\BibitemShut {NoStop}%
\bibitem [{\citenamefont {Staude}\ \emph {et~al.}(2013)\citenamefont {Staude},
  \citenamefont {Miroshnichenko}, \citenamefont {Decker}, \citenamefont
  {Fofang}, \citenamefont {Liu}, \citenamefont {Gonzales}, \citenamefont
  {Dominguez}, \citenamefont {Luk}, \citenamefont {Neshev}, \citenamefont
  {Brener} \emph {et~al.}}]{staude2013tailoring}%
  \BibitemOpen
  \bibfield  {author} {\bibinfo {author} {\bibfnamefont {I.}~\bibnamefont
  {Staude}}, \bibinfo {author} {\bibfnamefont {A.~E.}\ \bibnamefont
  {Miroshnichenko}}, \bibinfo {author} {\bibfnamefont {M.}~\bibnamefont
  {Decker}}, \bibinfo {author} {\bibfnamefont {N.~T.}\ \bibnamefont {Fofang}},
  \bibinfo {author} {\bibfnamefont {S.}~\bibnamefont {Liu}}, \bibinfo {author}
  {\bibfnamefont {E.}~\bibnamefont {Gonzales}}, \bibinfo {author}
  {\bibfnamefont {J.}~\bibnamefont {Dominguez}}, \bibinfo {author}
  {\bibfnamefont {T.~S.}\ \bibnamefont {Luk}}, \bibinfo {author} {\bibfnamefont
  {D.~N.}\ \bibnamefont {Neshev}}, \bibinfo {author} {\bibfnamefont
  {I.}~\bibnamefont {Brener}},  \emph {et~al.},\ }\href@noop {} {\bibfield
  {journal} {\bibinfo  {journal} {ACS Nano}\ }\textbf {\bibinfo {volume} {7}},\
  \bibinfo {pages} {7824} (\bibinfo {year} {2013})}\BibitemShut {NoStop}%
\bibitem [{\citenamefont {Nieto-Vesperinas}\ \emph {et~al.}(2011)\citenamefont
  {Nieto-Vesperinas}, \citenamefont {Gomez-Medina},\ and\ \citenamefont
  {Saenz}}]{nieto2011angle}%
  \BibitemOpen
  \bibfield  {author} {\bibinfo {author} {\bibfnamefont {M.}~\bibnamefont
  {Nieto-Vesperinas}}, \bibinfo {author} {\bibfnamefont {R.}~\bibnamefont
  {Gomez-Medina}}, \ and\ \bibinfo {author} {\bibfnamefont {J.}~\bibnamefont
  {Saenz}},\ }\href@noop {} {\bibfield  {journal} {\bibinfo  {journal} {JOSA
  A}\ }\textbf {\bibinfo {volume} {28}},\ \bibinfo {pages} {54} (\bibinfo
  {year} {2011})}\BibitemShut {NoStop}%
\bibitem [{\citenamefont {Liu}\ and\ \citenamefont
  {Kivshar}(2018)}]{liu2018generalized}%
  \BibitemOpen
  \bibfield  {author} {\bibinfo {author} {\bibfnamefont {W.}~\bibnamefont
  {Liu}}\ and\ \bibinfo {author} {\bibfnamefont {Y.~S.}\ \bibnamefont
  {Kivshar}},\ }\href@noop {} {\bibfield  {journal} {\bibinfo  {journal} {Opt.
  Express}\ }\textbf {\bibinfo {volume} {26}},\ \bibinfo {pages} {13085}
  (\bibinfo {year} {2018})}\BibitemShut {NoStop}%
\bibitem [{\citenamefont {Alaee}\ \emph {et~al.}(2015)\citenamefont {Alaee},
  \citenamefont {Filter}, \citenamefont {Lehr}, \citenamefont {Lederer},\ and\
  \citenamefont {Rockstuhl}}]{alaee2015generalized}%
  \BibitemOpen
  \bibfield  {author} {\bibinfo {author} {\bibfnamefont {R.}~\bibnamefont
  {Alaee}}, \bibinfo {author} {\bibfnamefont {R.}~\bibnamefont {Filter}},
  \bibinfo {author} {\bibfnamefont {D.}~\bibnamefont {Lehr}}, \bibinfo {author}
  {\bibfnamefont {F.}~\bibnamefont {Lederer}}, \ and\ \bibinfo {author}
  {\bibfnamefont {C.}~\bibnamefont {Rockstuhl}},\ }\href@noop {} {\bibfield
  {journal} {\bibinfo  {journal} {Opt. Lett.}\ }\textbf {\bibinfo {volume}
  {40}},\ \bibinfo {pages} {2645} (\bibinfo {year} {2015})}\BibitemShut
  {NoStop}%
\bibitem [{\citenamefont {Shamkhi}\ \emph {et~al.}(2019)\citenamefont
  {Shamkhi}, \citenamefont {Baryshnikova}, \citenamefont {Sayanskiy},
  \citenamefont {Kapitanova}, \citenamefont {Terekhov}, \citenamefont {Belov},
  \citenamefont {Karabchevsky}, \citenamefont {Evlyukhin}, \citenamefont
  {Kivshar},\ and\ \citenamefont {Shalin}}]{shamkhi2019transverse}%
  \BibitemOpen
  \bibfield  {author} {\bibinfo {author} {\bibfnamefont {H.~K.}\ \bibnamefont
  {Shamkhi}}, \bibinfo {author} {\bibfnamefont {K.~V.}\ \bibnamefont
  {Baryshnikova}}, \bibinfo {author} {\bibfnamefont {A.}~\bibnamefont
  {Sayanskiy}}, \bibinfo {author} {\bibfnamefont {P.}~\bibnamefont
  {Kapitanova}}, \bibinfo {author} {\bibfnamefont {P.~D.}\ \bibnamefont
  {Terekhov}}, \bibinfo {author} {\bibfnamefont {P.}~\bibnamefont {Belov}},
  \bibinfo {author} {\bibfnamefont {A.}~\bibnamefont {Karabchevsky}}, \bibinfo
  {author} {\bibfnamefont {A.~B.}\ \bibnamefont {Evlyukhin}}, \bibinfo {author}
  {\bibfnamefont {Y.}~\bibnamefont {Kivshar}}, \ and\ \bibinfo {author}
  {\bibfnamefont {A.~S.}\ \bibnamefont {Shalin}},\ }\href@noop {} {\bibfield
  {journal} {\bibinfo  {journal} {Phys. Rev. Lett.}\ }\textbf {\bibinfo
  {volume} {122}},\ \bibinfo {pages} {193905} (\bibinfo {year}
  {2019})}\BibitemShut {NoStop}%
\bibitem [{\citenamefont {Lu}\ \emph {et~al.}(2015)\citenamefont {Lu},
  \citenamefont {Wang}, \citenamefont {Chou}, \citenamefont {Shen},
  \citenamefont {He}, \citenamefont {Cheng},\ and\ \citenamefont
  {Gong}}]{lu2015directional}%
  \BibitemOpen
  \bibfield  {author} {\bibinfo {author} {\bibfnamefont {G.}~\bibnamefont
  {Lu}}, \bibinfo {author} {\bibfnamefont {Y.}~\bibnamefont {Wang}}, \bibinfo
  {author} {\bibfnamefont {R.~Y.}\ \bibnamefont {Chou}}, \bibinfo {author}
  {\bibfnamefont {H.}~\bibnamefont {Shen}}, \bibinfo {author} {\bibfnamefont
  {Y.}~\bibnamefont {He}}, \bibinfo {author} {\bibfnamefont {Y.}~\bibnamefont
  {Cheng}}, \ and\ \bibinfo {author} {\bibfnamefont {Q.}~\bibnamefont {Gong}},\
  }\href@noop {} {\bibfield  {journal} {\bibinfo  {journal} {Laser Photonics
  Rev.}\ }\textbf {\bibinfo {volume} {9}},\ \bibinfo {pages} {530} (\bibinfo
  {year} {2015})}\BibitemShut {NoStop}%
\bibitem [{\citenamefont {Sinev}\ \emph {et~al.}(2017)\citenamefont {Sinev},
  \citenamefont {Bogdanov}, \citenamefont {Komissarenko}, \citenamefont
  {Frizyuk}, \citenamefont {Petrov}, \citenamefont {Mukhin}, \citenamefont
  {Makarov}, \citenamefont {Samusev}, \citenamefont {Lavrinenko},\ and\
  \citenamefont {Iorsh}}]{sinev2017chirality}%
  \BibitemOpen
  \bibfield  {author} {\bibinfo {author} {\bibfnamefont {I.~S.}\ \bibnamefont
  {Sinev}}, \bibinfo {author} {\bibfnamefont {A.~A.}\ \bibnamefont {Bogdanov}},
  \bibinfo {author} {\bibfnamefont {F.~E.}\ \bibnamefont {Komissarenko}},
  \bibinfo {author} {\bibfnamefont {K.~S.}\ \bibnamefont {Frizyuk}}, \bibinfo
  {author} {\bibfnamefont {M.~I.}\ \bibnamefont {Petrov}}, \bibinfo {author}
  {\bibfnamefont {I.~S.}\ \bibnamefont {Mukhin}}, \bibinfo {author}
  {\bibfnamefont {S.~V.}\ \bibnamefont {Makarov}}, \bibinfo {author}
  {\bibfnamefont {A.~K.}\ \bibnamefont {Samusev}}, \bibinfo {author}
  {\bibfnamefont {A.~V.}\ \bibnamefont {Lavrinenko}}, \ and\ \bibinfo {author}
  {\bibfnamefont {I.~V.}\ \bibnamefont {Iorsh}},\ }\href@noop {} {\bibfield
  {journal} {\bibinfo  {journal} {Laser Photonics Rev.}\ }\textbf {\bibinfo
  {volume} {11}},\ \bibinfo {pages} {1700168} (\bibinfo {year}
  {2017})}\BibitemShut {NoStop}%
\bibitem [{\citenamefont {Sinev}\ \emph {et~al.}(2020)\citenamefont {Sinev},
  \citenamefont {Komissarenko}, \citenamefont {Iorsh}, \citenamefont
  {Permyakov}, \citenamefont {Samusev},\ and\ \citenamefont
  {Bogdanov}}]{sinev2020steering}%
  \BibitemOpen
  \bibfield  {author} {\bibinfo {author} {\bibfnamefont {I.}~\bibnamefont
  {Sinev}}, \bibinfo {author} {\bibfnamefont {F.}~\bibnamefont {Komissarenko}},
  \bibinfo {author} {\bibfnamefont {I.}~\bibnamefont {Iorsh}}, \bibinfo
  {author} {\bibfnamefont {D.}~\bibnamefont {Permyakov}}, \bibinfo {author}
  {\bibfnamefont {A.}~\bibnamefont {Samusev}}, \ and\ \bibinfo {author}
  {\bibfnamefont {A.}~\bibnamefont {Bogdanov}},\ }\href@noop {} {\bibfield
  {journal} {\bibinfo  {journal} {ACS Photonics}\ }\textbf {\bibinfo {volume}
  {7}},\ \bibinfo {pages} {680} (\bibinfo {year} {2020})}\BibitemShut {NoStop}%
\bibitem [{\citenamefont {Shcherbakov}\ \emph {et~al.}(2014)\citenamefont
  {Shcherbakov}, \citenamefont {Neshev}, \citenamefont {Hopkins}, \citenamefont
  {Shorokhov}, \citenamefont {Staude}, \citenamefont {Melik-Gaykazyan},
  \citenamefont {Decker}, \citenamefont {Ezhov}, \citenamefont
  {Miroshnichenko}, \citenamefont {Brener} \emph
  {et~al.}}]{shcherbakov2014enhanced}%
  \BibitemOpen
  \bibfield  {author} {\bibinfo {author} {\bibfnamefont {M.~R.}\ \bibnamefont
  {Shcherbakov}}, \bibinfo {author} {\bibfnamefont {D.~N.}\ \bibnamefont
  {Neshev}}, \bibinfo {author} {\bibfnamefont {B.}~\bibnamefont {Hopkins}},
  \bibinfo {author} {\bibfnamefont {A.~S.}\ \bibnamefont {Shorokhov}}, \bibinfo
  {author} {\bibfnamefont {I.}~\bibnamefont {Staude}}, \bibinfo {author}
  {\bibfnamefont {E.~V.}\ \bibnamefont {Melik-Gaykazyan}}, \bibinfo {author}
  {\bibfnamefont {M.}~\bibnamefont {Decker}}, \bibinfo {author} {\bibfnamefont
  {A.~A.}\ \bibnamefont {Ezhov}}, \bibinfo {author} {\bibfnamefont {A.~E.}\
  \bibnamefont {Miroshnichenko}}, \bibinfo {author} {\bibfnamefont
  {I.}~\bibnamefont {Brener}},  \emph {et~al.},\ }\href@noop {} {\bibfield
  {journal} {\bibinfo  {journal} {Nano Lett.}\ }\textbf {\bibinfo {volume}
  {14}},\ \bibinfo {pages} {6488} (\bibinfo {year} {2014})}\BibitemShut
  {NoStop}%
\bibitem [{\citenamefont {Makarov}\ \emph {et~al.}(2015)\citenamefont
  {Makarov}, \citenamefont {Kudryashov}, \citenamefont {Mukhin}, \citenamefont
  {Mozharov}, \citenamefont {Milichko}, \citenamefont {Krasnok},\ and\
  \citenamefont {Belov}}]{makarov2015tuning}%
  \BibitemOpen
  \bibfield  {author} {\bibinfo {author} {\bibfnamefont {S.}~\bibnamefont
  {Makarov}}, \bibinfo {author} {\bibfnamefont {S.}~\bibnamefont {Kudryashov}},
  \bibinfo {author} {\bibfnamefont {I.}~\bibnamefont {Mukhin}}, \bibinfo
  {author} {\bibfnamefont {A.}~\bibnamefont {Mozharov}}, \bibinfo {author}
  {\bibfnamefont {V.}~\bibnamefont {Milichko}}, \bibinfo {author}
  {\bibfnamefont {A.}~\bibnamefont {Krasnok}}, \ and\ \bibinfo {author}
  {\bibfnamefont {P.}~\bibnamefont {Belov}},\ }\href@noop {} {\bibfield
  {journal} {\bibinfo  {journal} {Nano Lett.}\ }\textbf {\bibinfo {volume}
  {15}},\ \bibinfo {pages} {6187} (\bibinfo {year} {2015})}\BibitemShut
  {NoStop}%
\bibitem [{\citenamefont {Shcherbakov}\ \emph {et~al.}(2015)\citenamefont
  {Shcherbakov}, \citenamefont {Vabishchevich}, \citenamefont {Shorokhov},
  \citenamefont {Chong}, \citenamefont {Choi}, \citenamefont {Staude},
  \citenamefont {Miroshnichenko}, \citenamefont {Neshev}, \citenamefont
  {Fedyanin},\ and\ \citenamefont {Kivshar}}]{shcherbakov2015ultrafast}%
  \BibitemOpen
  \bibfield  {author} {\bibinfo {author} {\bibfnamefont {M.~R.}\ \bibnamefont
  {Shcherbakov}}, \bibinfo {author} {\bibfnamefont {P.~P.}\ \bibnamefont
  {Vabishchevich}}, \bibinfo {author} {\bibfnamefont {A.~S.}\ \bibnamefont
  {Shorokhov}}, \bibinfo {author} {\bibfnamefont {K.~E.}\ \bibnamefont
  {Chong}}, \bibinfo {author} {\bibfnamefont {D.-Y.}\ \bibnamefont {Choi}},
  \bibinfo {author} {\bibfnamefont {I.}~\bibnamefont {Staude}}, \bibinfo
  {author} {\bibfnamefont {A.~E.}\ \bibnamefont {Miroshnichenko}}, \bibinfo
  {author} {\bibfnamefont {D.~N.}\ \bibnamefont {Neshev}}, \bibinfo {author}
  {\bibfnamefont {A.~A.}\ \bibnamefont {Fedyanin}}, \ and\ \bibinfo {author}
  {\bibfnamefont {Y.~S.}\ \bibnamefont {Kivshar}},\ }\href@noop {} {\bibfield
  {journal} {\bibinfo  {journal} {Nano Lett.}\ }\textbf {\bibinfo {volume}
  {15}},\ \bibinfo {pages} {6985} (\bibinfo {year} {2015})}\BibitemShut
  {NoStop}%
\bibitem [{\citenamefont {Carletti}\ \emph {et~al.}(2019)\citenamefont
  {Carletti}, \citenamefont {Kruk}, \citenamefont {Bogdanov}, \citenamefont
  {De~Angelis},\ and\ \citenamefont {Kivshar}}]{carletti2019high}%
  \BibitemOpen
  \bibfield  {author} {\bibinfo {author} {\bibfnamefont {L.}~\bibnamefont
  {Carletti}}, \bibinfo {author} {\bibfnamefont {S.~S.}\ \bibnamefont {Kruk}},
  \bibinfo {author} {\bibfnamefont {A.~A.}\ \bibnamefont {Bogdanov}}, \bibinfo
  {author} {\bibfnamefont {C.}~\bibnamefont {De~Angelis}}, \ and\ \bibinfo
  {author} {\bibfnamefont {Y.}~\bibnamefont {Kivshar}},\ }\href@noop {}
  {\bibfield  {journal} {\bibinfo  {journal} {Phys. Rev. Research}\ }\textbf
  {\bibinfo {volume} {1}},\ \bibinfo {pages} {023016} (\bibinfo {year}
  {2019})}\BibitemShut {NoStop}%
\bibitem [{\citenamefont {Koshelev}\ \emph {et~al.}(2020)\citenamefont
  {Koshelev}, \citenamefont {Kruk}, \citenamefont {Melik-Gaykazyan},
  \citenamefont {Choi}, \citenamefont {Bogdanov}, \citenamefont {Park},\ and\
  \citenamefont {Kivshar}}]{koshelev2020subwavelength}%
  \BibitemOpen
  \bibfield  {author} {\bibinfo {author} {\bibfnamefont {K.}~\bibnamefont
  {Koshelev}}, \bibinfo {author} {\bibfnamefont {S.}~\bibnamefont {Kruk}},
  \bibinfo {author} {\bibfnamefont {E.}~\bibnamefont {Melik-Gaykazyan}},
  \bibinfo {author} {\bibfnamefont {J.-H.}\ \bibnamefont {Choi}}, \bibinfo
  {author} {\bibfnamefont {A.}~\bibnamefont {Bogdanov}}, \bibinfo {author}
  {\bibfnamefont {H.-G.}\ \bibnamefont {Park}}, \ and\ \bibinfo {author}
  {\bibfnamefont {Y.}~\bibnamefont {Kivshar}},\ }\href@noop {} {\bibfield
  {journal} {\bibinfo  {journal} {Science}\ }\textbf {\bibinfo {volume}
  {367}},\ \bibinfo {pages} {288} (\bibinfo {year} {2020})}\BibitemShut
  {NoStop}%
\bibitem [{\citenamefont {Tiguntseva}\ \emph {et~al.}(2020)\citenamefont
  {Tiguntseva}, \citenamefont {Koshelev}, \citenamefont {Furasova},
  \citenamefont {Tonkaev}, \citenamefont {Mikhailovskii}, \citenamefont
  {Ushakova}, \citenamefont {Baranov}, \citenamefont {Shegai}, \citenamefont
  {Zakhidov}, \citenamefont {Kivshar} \emph {et~al.}}]{tiguntseva2020room}%
  \BibitemOpen
  \bibfield  {author} {\bibinfo {author} {\bibfnamefont {E.}~\bibnamefont
  {Tiguntseva}}, \bibinfo {author} {\bibfnamefont {K.}~\bibnamefont
  {Koshelev}}, \bibinfo {author} {\bibfnamefont {A.}~\bibnamefont {Furasova}},
  \bibinfo {author} {\bibfnamefont {P.}~\bibnamefont {Tonkaev}}, \bibinfo
  {author} {\bibfnamefont {V.}~\bibnamefont {Mikhailovskii}}, \bibinfo {author}
  {\bibfnamefont {E.~V.}\ \bibnamefont {Ushakova}}, \bibinfo {author}
  {\bibfnamefont {D.~G.}\ \bibnamefont {Baranov}}, \bibinfo {author}
  {\bibfnamefont {T.}~\bibnamefont {Shegai}}, \bibinfo {author} {\bibfnamefont
  {A.~A.}\ \bibnamefont {Zakhidov}}, \bibinfo {author} {\bibfnamefont
  {Y.}~\bibnamefont {Kivshar}},  \emph {et~al.},\ }\href@noop {} {\bibfield
  {journal} {\bibinfo  {journal} {ACS Nano}\ } (\bibinfo {year}
  {2020})}\BibitemShut {NoStop}%
\bibitem [{\citenamefont {Bohren}\ and\ \citenamefont
  {Huffman}(2008)}]{bohren2008absorption}%
  \BibitemOpen
  \bibfield  {author} {\bibinfo {author} {\bibfnamefont {C.~F.}\ \bibnamefont
  {Bohren}}\ and\ \bibinfo {author} {\bibfnamefont {D.~R.}\ \bibnamefont
  {Huffman}},\ }\href@noop {} {\emph {\bibinfo {title} {Absorption and
  scattering of light by small particles}}}\ (\bibinfo  {publisher} {John Wiley
  \& Sons},\ \bibinfo {year} {2008})\BibitemShut {NoStop}%
\bibitem [{\citenamefont {Kerker}(2013)}]{kerker2013scattering}%
  \BibitemOpen
  \bibfield  {author} {\bibinfo {author} {\bibfnamefont {M.}~\bibnamefont
  {Kerker}},\ }\href@noop {} {\emph {\bibinfo {title} {The scattering of light
  and other electromagnetic radiation: physical chemistry: a series of
  monographs}}},\ Vol.~\bibinfo {volume} {16}\ (\bibinfo  {publisher} {Academic
  Press},\ \bibinfo {year} {2013})\BibitemShut {NoStop}%
\bibitem [{\citenamefont {Van~de Groep}\ and\ \citenamefont
  {Polman}(2013)}]{van2013designing}%
  \BibitemOpen
  \bibfield  {author} {\bibinfo {author} {\bibfnamefont {J.}~\bibnamefont
  {Van~de Groep}}\ and\ \bibinfo {author} {\bibfnamefont {A.}~\bibnamefont
  {Polman}},\ }\href@noop {} {\bibfield  {journal} {\bibinfo  {journal} {Opt.
  Express}\ }\textbf {\bibinfo {volume} {21}},\ \bibinfo {pages} {26285}
  (\bibinfo {year} {2013})}\BibitemShut {NoStop}%
\bibitem [{\citenamefont {Carletti}\ \emph {et~al.}(2015)\citenamefont
  {Carletti}, \citenamefont {Locatelli}, \citenamefont {Stepanenko},
  \citenamefont {Leo},\ and\ \citenamefont
  {De~Angelis}}]{carletti2015enhanced}%
  \BibitemOpen
  \bibfield  {author} {\bibinfo {author} {\bibfnamefont {L.}~\bibnamefont
  {Carletti}}, \bibinfo {author} {\bibfnamefont {A.}~\bibnamefont {Locatelli}},
  \bibinfo {author} {\bibfnamefont {O.}~\bibnamefont {Stepanenko}}, \bibinfo
  {author} {\bibfnamefont {G.}~\bibnamefont {Leo}}, \ and\ \bibinfo {author}
  {\bibfnamefont {C.}~\bibnamefont {De~Angelis}},\ }\href@noop {} {\bibfield
  {journal} {\bibinfo  {journal} {Opt. Express}\ }\textbf {\bibinfo {volume}
  {23}},\ \bibinfo {pages} {26544} (\bibinfo {year} {2015})}\BibitemShut
  {NoStop}%
\bibitem [{\citenamefont {Melik-Gaykazyan}\ \emph {et~al.}(2017)\citenamefont
  {Melik-Gaykazyan}, \citenamefont {Kruk}, \citenamefont {Camacho-Morales},
  \citenamefont {Xu}, \citenamefont {Rahmani}, \citenamefont {Zangeneh~Kamali},
  \citenamefont {Lamprianidis}, \citenamefont {Miroshnichenko}, \citenamefont
  {Fedyanin}, \citenamefont {Neshev} \emph {et~al.}}]{melik2017selective}%
  \BibitemOpen
  \bibfield  {author} {\bibinfo {author} {\bibfnamefont {E.~V.}\ \bibnamefont
  {Melik-Gaykazyan}}, \bibinfo {author} {\bibfnamefont {S.~S.}\ \bibnamefont
  {Kruk}}, \bibinfo {author} {\bibfnamefont {R.}~\bibnamefont
  {Camacho-Morales}}, \bibinfo {author} {\bibfnamefont {L.}~\bibnamefont {Xu}},
  \bibinfo {author} {\bibfnamefont {M.}~\bibnamefont {Rahmani}}, \bibinfo
  {author} {\bibfnamefont {K.}~\bibnamefont {Zangeneh~Kamali}}, \bibinfo
  {author} {\bibfnamefont {A.}~\bibnamefont {Lamprianidis}}, \bibinfo {author}
  {\bibfnamefont {A.~E.}\ \bibnamefont {Miroshnichenko}}, \bibinfo {author}
  {\bibfnamefont {A.~A.}\ \bibnamefont {Fedyanin}}, \bibinfo {author}
  {\bibfnamefont {D.~N.}\ \bibnamefont {Neshev}},  \emph {et~al.},\ }\href@noop
  {} {\bibfield  {journal} {\bibinfo  {journal} {ACS Photonics}\ }\textbf
  {\bibinfo {volume} {5}},\ \bibinfo {pages} {728} (\bibinfo {year}
  {2017})}\BibitemShut {NoStop}%
\bibitem [{\citenamefont {Liu}\ \emph {et~al.}(2017)\citenamefont {Liu},
  \citenamefont {Panmai}, \citenamefont {Peng},\ and\ \citenamefont
  {Lan}}]{liu2017optical}%
  \BibitemOpen
  \bibfield  {author} {\bibinfo {author} {\bibfnamefont {H.}~\bibnamefont
  {Liu}}, \bibinfo {author} {\bibfnamefont {M.}~\bibnamefont {Panmai}},
  \bibinfo {author} {\bibfnamefont {Y.}~\bibnamefont {Peng}}, \ and\ \bibinfo
  {author} {\bibfnamefont {S.}~\bibnamefont {Lan}},\ }\href@noop {} {\bibfield
  {journal} {\bibinfo  {journal} {Opt. Express}\ }\textbf {\bibinfo {volume}
  {25}},\ \bibinfo {pages} {12357} (\bibinfo {year} {2017})}\BibitemShut
  {NoStop}%
\bibitem [{\citenamefont {Nagasaki}\ \emph {et~al.}(2017)\citenamefont
  {Nagasaki}, \citenamefont {Suzuki},\ and\ \citenamefont
  {Takahara}}]{nagasaki2017all}%
  \BibitemOpen
  \bibfield  {author} {\bibinfo {author} {\bibfnamefont {Y.}~\bibnamefont
  {Nagasaki}}, \bibinfo {author} {\bibfnamefont {M.}~\bibnamefont {Suzuki}}, \
  and\ \bibinfo {author} {\bibfnamefont {J.}~\bibnamefont {Takahara}},\
  }\href@noop {} {\bibfield  {journal} {\bibinfo  {journal} {Nano Lett.}\
  }\textbf {\bibinfo {volume} {17}},\ \bibinfo {pages} {7500} (\bibinfo {year}
  {2017})}\BibitemShut {NoStop}%
\bibitem [{\citenamefont {Flauraud}\ \emph {et~al.}(2017)\citenamefont
  {Flauraud}, \citenamefont {Reyes}, \citenamefont {Paniagua-Dominguez},
  \citenamefont {Kuznetsov},\ and\ \citenamefont
  {Brugger}}]{flauraud2017silicon}%
  \BibitemOpen
  \bibfield  {author} {\bibinfo {author} {\bibfnamefont {V.}~\bibnamefont
  {Flauraud}}, \bibinfo {author} {\bibfnamefont {M.}~\bibnamefont {Reyes}},
  \bibinfo {author} {\bibfnamefont {R.}~\bibnamefont {Paniagua-Dominguez}},
  \bibinfo {author} {\bibfnamefont {A.~I.}\ \bibnamefont {Kuznetsov}}, \ and\
  \bibinfo {author} {\bibfnamefont {J.}~\bibnamefont {Brugger}},\ }\href@noop
  {} {\bibfield  {journal} {\bibinfo  {journal} {ACS Photonics}\ }\textbf
  {\bibinfo {volume} {4}},\ \bibinfo {pages} {1913} (\bibinfo {year}
  {2017})}\BibitemShut {NoStop}%
\bibitem [{\citenamefont {Xiang}\ \emph {et~al.}(2018)\citenamefont {Xiang},
  \citenamefont {Li}, \citenamefont {Zhou}, \citenamefont {Jiang},
  \citenamefont {Chen}, \citenamefont {Dai}, \citenamefont {Tie}, \citenamefont
  {Lan},\ and\ \citenamefont {Wang}}]{xiang2018manipulating}%
  \BibitemOpen
  \bibfield  {author} {\bibinfo {author} {\bibfnamefont {J.}~\bibnamefont
  {Xiang}}, \bibinfo {author} {\bibfnamefont {J.}~\bibnamefont {Li}}, \bibinfo
  {author} {\bibfnamefont {Z.}~\bibnamefont {Zhou}}, \bibinfo {author}
  {\bibfnamefont {S.}~\bibnamefont {Jiang}}, \bibinfo {author} {\bibfnamefont
  {J.}~\bibnamefont {Chen}}, \bibinfo {author} {\bibfnamefont {Q.}~\bibnamefont
  {Dai}}, \bibinfo {author} {\bibfnamefont {S.}~\bibnamefont {Tie}}, \bibinfo
  {author} {\bibfnamefont {S.}~\bibnamefont {Lan}}, \ and\ \bibinfo {author}
  {\bibfnamefont {X.}~\bibnamefont {Wang}},\ }\href@noop {} {\bibfield
  {journal} {\bibinfo  {journal} {Laser Photonics Rev.}\ }\textbf {\bibinfo
  {volume} {12}},\ \bibinfo {pages} {1800032} (\bibinfo {year}
  {2018})}\BibitemShut {NoStop}%
\bibitem [{\citenamefont {Li}\ \emph {et~al.}(2019)\citenamefont {Li},
  \citenamefont {Xu}, \citenamefont {Veetil}, \citenamefont {Valuckas},
  \citenamefont {Paniagua-Dom{\'\i}nguez},\ and\ \citenamefont
  {Kuznetsov}}]{li2019phase}%
  \BibitemOpen
  \bibfield  {author} {\bibinfo {author} {\bibfnamefont {S.-Q.}\ \bibnamefont
  {Li}}, \bibinfo {author} {\bibfnamefont {X.}~\bibnamefont {Xu}}, \bibinfo
  {author} {\bibfnamefont {R.~M.}\ \bibnamefont {Veetil}}, \bibinfo {author}
  {\bibfnamefont {V.}~\bibnamefont {Valuckas}}, \bibinfo {author}
  {\bibfnamefont {R.}~\bibnamefont {Paniagua-Dom{\'\i}nguez}}, \ and\ \bibinfo
  {author} {\bibfnamefont {A.~I.}\ \bibnamefont {Kuznetsov}},\ }\href@noop {}
  {\bibfield  {journal} {\bibinfo  {journal} {Science}\ }\textbf {\bibinfo
  {volume} {364}},\ \bibinfo {pages} {1087} (\bibinfo {year}
  {2019})}\BibitemShut {NoStop}%
\bibitem [{\citenamefont {Wo{\'z}niak}\ \emph {et~al.}(2015)\citenamefont
  {Wo{\'z}niak}, \citenamefont {Banzer},\ and\ \citenamefont
  {Leuchs}}]{wozniak2015selective}%
  \BibitemOpen
  \bibfield  {author} {\bibinfo {author} {\bibfnamefont {P.}~\bibnamefont
  {Wo{\'z}niak}}, \bibinfo {author} {\bibfnamefont {P.}~\bibnamefont {Banzer}},
  \ and\ \bibinfo {author} {\bibfnamefont {G.}~\bibnamefont {Leuchs}},\
  }\href@noop {} {\bibfield  {journal} {\bibinfo  {journal} {Laser Photonics
  Rev.}\ }\textbf {\bibinfo {volume} {9}},\ \bibinfo {pages} {231} (\bibinfo
  {year} {2015})}\BibitemShut {NoStop}%
\bibitem [{\citenamefont {Neugebauer}\ \emph {et~al.}(2016)\citenamefont
  {Neugebauer}, \citenamefont {Wo{\'z}niak}, \citenamefont {Bag}, \citenamefont
  {Leuchs},\ and\ \citenamefont {Banzer}}]{neugebauer2016polarization}%
  \BibitemOpen
  \bibfield  {author} {\bibinfo {author} {\bibfnamefont {M.}~\bibnamefont
  {Neugebauer}}, \bibinfo {author} {\bibfnamefont {P.}~\bibnamefont
  {Wo{\'z}niak}}, \bibinfo {author} {\bibfnamefont {A.}~\bibnamefont {Bag}},
  \bibinfo {author} {\bibfnamefont {G.}~\bibnamefont {Leuchs}}, \ and\ \bibinfo
  {author} {\bibfnamefont {P.}~\bibnamefont {Banzer}},\ }\href@noop {}
  {\bibfield  {journal} {\bibinfo  {journal} {Nat. Commun.}\ }\textbf {\bibinfo
  {volume} {7}},\ \bibinfo {pages} {1} (\bibinfo {year} {2016})}\BibitemShut
  {NoStop}%
\bibitem [{\citenamefont {Sinev}\ \emph {et~al.}(2016)\citenamefont {Sinev},
  \citenamefont {Iorsh}, \citenamefont {Bogdanov}, \citenamefont {Permyakov},
  \citenamefont {Komissarenko}, \citenamefont {Mukhin}, \citenamefont
  {Samusev}, \citenamefont {Valuckas}, \citenamefont {Kuznetsov}, \citenamefont
  {Luk'yanchuk} \emph {et~al.}}]{sinev2016polarization}%
  \BibitemOpen
  \bibfield  {author} {\bibinfo {author} {\bibfnamefont {I.}~\bibnamefont
  {Sinev}}, \bibinfo {author} {\bibfnamefont {I.}~\bibnamefont {Iorsh}},
  \bibinfo {author} {\bibfnamefont {A.}~\bibnamefont {Bogdanov}}, \bibinfo
  {author} {\bibfnamefont {D.}~\bibnamefont {Permyakov}}, \bibinfo {author}
  {\bibfnamefont {F.}~\bibnamefont {Komissarenko}}, \bibinfo {author}
  {\bibfnamefont {I.}~\bibnamefont {Mukhin}}, \bibinfo {author} {\bibfnamefont
  {A.}~\bibnamefont {Samusev}}, \bibinfo {author} {\bibfnamefont
  {V.}~\bibnamefont {Valuckas}}, \bibinfo {author} {\bibfnamefont {A.~I.}\
  \bibnamefont {Kuznetsov}}, \bibinfo {author} {\bibfnamefont {B.~S.}\
  \bibnamefont {Luk'yanchuk}},  \emph {et~al.},\ }\href@noop {} {\bibfield
  {journal} {\bibinfo  {journal} {Laser Photonics Rev.}\ }\textbf {\bibinfo
  {volume} {10}},\ \bibinfo {pages} {799} (\bibinfo {year} {2016})}\BibitemShut
  {NoStop}%
\bibitem [{\citenamefont {Gladyshev}\ \emph {et~al.}(2020)\citenamefont
  {Gladyshev}, \citenamefont {Frizyuk},\ and\ \citenamefont
  {Bogdanov}}]{gladyshev2020symmetry}%
  \BibitemOpen
  \bibfield  {author} {\bibinfo {author} {\bibfnamefont {S.}~\bibnamefont
  {Gladyshev}}, \bibinfo {author} {\bibfnamefont {K.}~\bibnamefont {Frizyuk}},
  \ and\ \bibinfo {author} {\bibfnamefont {A.}~\bibnamefont {Bogdanov}},\
  }\href@noop {} {\bibfield  {journal} {\bibinfo  {journal} {Phys. Rev. B}\
  }\textbf {\bibinfo {volume} {102}},\ \bibinfo {pages} {075103} (\bibinfo
  {year} {2020})}\BibitemShut {NoStop}%
\bibitem [{\citenamefont {Arbabi}\ \emph {et~al.}(2017)\citenamefont {Arbabi},
  \citenamefont {Arbabi}, \citenamefont {Horie}, \citenamefont {Kamali},\ and\
  \citenamefont {Faraon}}]{arbabi2017planar}%
  \BibitemOpen
  \bibfield  {author} {\bibinfo {author} {\bibfnamefont {A.}~\bibnamefont
  {Arbabi}}, \bibinfo {author} {\bibfnamefont {E.}~\bibnamefont {Arbabi}},
  \bibinfo {author} {\bibfnamefont {Y.}~\bibnamefont {Horie}}, \bibinfo
  {author} {\bibfnamefont {S.~M.}\ \bibnamefont {Kamali}}, \ and\ \bibinfo
  {author} {\bibfnamefont {A.}~\bibnamefont {Faraon}},\ }\href@noop {}
  {\bibfield  {journal} {\bibinfo  {journal} {Nat. Photonics}\ }\textbf
  {\bibinfo {volume} {11}},\ \bibinfo {pages} {415} (\bibinfo {year}
  {2017})}\BibitemShut {NoStop}%
\bibitem [{\citenamefont {Egel}\ \emph
  {et~al.}(2016{\natexlab{a}})\citenamefont {Egel}, \citenamefont {Kettlitz},\
  and\ \citenamefont {Lemmer}}]{egel2016efficient}%
  \BibitemOpen
  \bibfield  {author} {\bibinfo {author} {\bibfnamefont {A.}~\bibnamefont
  {Egel}}, \bibinfo {author} {\bibfnamefont {S.~W.}\ \bibnamefont {Kettlitz}},
  \ and\ \bibinfo {author} {\bibfnamefont {U.}~\bibnamefont {Lemmer}},\
  }\href@noop {} {\bibfield  {journal} {\bibinfo  {journal} {JOSA A}\ }\textbf
  {\bibinfo {volume} {33}},\ \bibinfo {pages} {698} (\bibinfo {year}
  {2016}{\natexlab{a}})}\BibitemShut {NoStop}%
\bibitem [{\citenamefont {Egel}\ \emph
  {et~al.}(2016{\natexlab{b}})\citenamefont {Egel}, \citenamefont {Theobald},
  \citenamefont {Donie}, \citenamefont {Lemmer},\ and\ \citenamefont
  {Gomard}}]{egel2016light}%
  \BibitemOpen
  \bibfield  {author} {\bibinfo {author} {\bibfnamefont {A.}~\bibnamefont
  {Egel}}, \bibinfo {author} {\bibfnamefont {D.}~\bibnamefont {Theobald}},
  \bibinfo {author} {\bibfnamefont {Y.}~\bibnamefont {Donie}}, \bibinfo
  {author} {\bibfnamefont {U.}~\bibnamefont {Lemmer}}, \ and\ \bibinfo {author}
  {\bibfnamefont {G.}~\bibnamefont {Gomard}},\ }\href@noop {} {\bibfield
  {journal} {\bibinfo  {journal} {Opt. Express}\ }\textbf {\bibinfo {volume}
  {24}},\ \bibinfo {pages} {25154} (\bibinfo {year}
  {2016}{\natexlab{b}})}\BibitemShut {NoStop}%
\bibitem [{\citenamefont {Egel}\ \emph {et~al.}(2017)\citenamefont {Egel},
  \citenamefont {Eremin}, \citenamefont {Wriedt}, \citenamefont {Theobald},
  \citenamefont {Lemmer},\ and\ \citenamefont {Gomard}}]{egel2017extending}%
  \BibitemOpen
  \bibfield  {author} {\bibinfo {author} {\bibfnamefont {A.}~\bibnamefont
  {Egel}}, \bibinfo {author} {\bibfnamefont {Y.}~\bibnamefont {Eremin}},
  \bibinfo {author} {\bibfnamefont {T.}~\bibnamefont {Wriedt}}, \bibinfo
  {author} {\bibfnamefont {D.}~\bibnamefont {Theobald}}, \bibinfo {author}
  {\bibfnamefont {U.}~\bibnamefont {Lemmer}}, \ and\ \bibinfo {author}
  {\bibfnamefont {G.}~\bibnamefont {Gomard}},\ }\href@noop {} {\bibfield
  {journal} {\bibinfo  {journal} {J. Quant. Spectrosc. Radiat. Transfer}\
  }\textbf {\bibinfo {volume} {202}},\ \bibinfo {pages} {279} (\bibinfo {year}
  {2017})}\BibitemShut {NoStop}%
\bibitem [{\citenamefont {Miroshnichenko}\ \emph {et~al.}(2015)\citenamefont
  {Miroshnichenko}, \citenamefont {Evlyukhin}, \citenamefont {Kivshar},\ and\
  \citenamefont {Chichkov}}]{miroshnichenko2015substrate}%
  \BibitemOpen
  \bibfield  {author} {\bibinfo {author} {\bibfnamefont {A.~E.}\ \bibnamefont
  {Miroshnichenko}}, \bibinfo {author} {\bibfnamefont {A.~B.}\ \bibnamefont
  {Evlyukhin}}, \bibinfo {author} {\bibfnamefont {Y.~S.}\ \bibnamefont
  {Kivshar}}, \ and\ \bibinfo {author} {\bibfnamefont {B.~N.}\ \bibnamefont
  {Chichkov}},\ }\href@noop {} {\bibfield  {journal} {\bibinfo  {journal} {ACS
  Photonics}\ }\textbf {\bibinfo {volume} {2}},\ \bibinfo {pages} {1423}
  (\bibinfo {year} {2015})}\BibitemShut {NoStop}%
\end{thebibliography}%

\end{document}